\documentclass[pdftex,twocolumn,epjc3]{svjour3}          

\RequirePackage[T1]{fontenc}

\smartqed  

\RequirePackage{graphicx}
\RequirePackage{newtxtext,newtxmath}      
\RequirePackage{flushend}
\RequirePackage[numbers,sort&compress]{natbib}
\RequirePackage[colorlinks,citecolor=blue,urlcolor=blue,linkcolor=blue]{hyperref}
\usepackage{xspace}
\usepackage[dvipsnames]{xcolor}

\newcommand{\nue}{\ensuremath{\nu_{e}}\xspace}
\newcommand{\numu}{\ensuremath{\nu_{\mu}}\xspace}
\newcommand{\nutau}{\ensuremath{\nu_{\tau}}\xspace}

\newcommand{\greco}{\ensuremath{\mathcal{A}}\xspace} 
\newcommand{\dragon}{\ensuremath{\mathcal{B}}\xspace}

\journalname{Eur. Phys. J. C}

\usepackage{enumitem}
\usepackage{booktabs}
\usepackage[referable]{threeparttablex}
\usepackage{todonotes}
\usepackage{comment}
\renewlist{tablenotes}{enumerate}{1}
\makeatletter
\setlist[tablenotes]{label=\tnote{\alph*},ref=\alph*,itemsep=\z@,topsep=\z@skip,partopsep=\z@skip,parsep=\z@,itemindent=\z@,labelindent=\tabcolsep,labelsep=.2em,leftmargin=*,align=left,before={\footnotesize}}
\makeatother

\newcommand{\newtext}[1]{\textcolor{black}{#1}\xspace}
\newcommand{\newnewtext}[1]{\textcolor{black}{#1}\xspace}
\newcommand{\newcolour}{\color{black}}
\newcommand{\oldcolour}{\color{black}}

\usepackage[tight]{subfigure}
\usepackage[symbol]{footmisc}

\begin{document}

\title{\newtext{Development of an analysis to probe the neutrino mass ordering with atmospheric neutrinos using three years of IceCube DeepCore data}}


\onecolumn

\author{IceCube Collaboration\footnote[1]{analysis@icecube.wisc.edu}: M.~G.~Aartsen\thanksref{Christchurch}
\and M.~Ackermann\thanksref{Zeuthen}
\and J.~Adams\thanksref{Christchurch}
\and J.~A.~Aguilar\thanksref{BrusselsLibre}
\and M.~Ahlers\thanksref{Copenhagen}
\and M.~Ahrens\thanksref{StockholmOKC}
\and C.~Alispach\thanksref{Geneva}
\and K.~Andeen\thanksref{Marquette}
\and T.~Anderson\thanksref{PennPhys}
\and I.~Ansseau\thanksref{BrusselsLibre}
% ...
}
\authorrunning{IceCube Collaboration}
\thankstext{a}{Earthquake Research Institute, University of Tokyo, Bunkyo, Tokyo 113-0032, Japan}
\institute{
% ...
Universit\'e Libre de Bruxelles, Science Faculty CP230, B-1050 Brussels, Belgium \label{BrusselsLibre}
% ...
\and Dept.~of Physics and Astronomy, University of Canterbury, Private Bag 4800, Christchurch, New Zealand \label{Christchurch}
% ...
\and Niels Bohr Institute, University of Copenhagen, DK-2100 Copenhagen, Denmark \label{Copenhagen}
% ...
\and D\'epartement de physique nucl\'eaire et corpusculaire, Universit\'e de Gen\`eve, CH-1211 Gen\`eve, Switzerland \label{Geneva}
% ...
\and Department of Physics, Marquette University, Milwaukee, WI, 53201, USA \label{Marquette}
% ...
\and Oskar Klein Centre and Dept.~of Physics, Stockholm University, SE-10691 Stockholm, Sweden \label{StockholmOKC}
% ...
\and Dept.~of Physics, Pennsylvania State University, University Park, PA 16802, USA \label{PennPhys}
% ...
\and DESY, D-15738 Zeuthen, Germany \label{Zeuthen}
} 
\date{Received: date / Accepted: date}
\maketitle
\twocolumn

\begin{abstract}
The Neutrino Mass Ordering (NMO) remains one of the outstanding questions in the field of neutrino physics. One strategy to measure the NMO is to observe matter effects in the oscillation pattern of atmospheric neutrinos above $\sim 1\,\mathrm{GeV}$, as proposed for several next-generation neutrino experiments.
Moreover, the existing IceCube DeepCore detector can already explore this type of measurement. We present \newtext{the development and application of two independent analyses to search for the signature of the NMO with three years of DeepCore data.}
These analyses include a full treatment of systematic uncertainties and a statistically-rigorous method to determine the significance for the NMO from a fit to the data.
\newtext{Both analyses show that the dataset is fully compatible with both mass orderings.}
For the more sensitive analysis, we observe a preference for normal ordering with a $p$-value of $p_\mathrm{IO} = 15.3\%$ and $\mathrm{CL}_\mathrm{s}=53.3\%$ for the inverted ordering hypothesis, while the experimental results from both analyses are consistent within their uncertainties. Since the result is independent of the value of $\delta_\mathrm{CP}$ and obtained from energies $E_\nu \gtrsim 5\,\mathrm{GeV}$, it is complementary to recent results from long-baseline experiments. These analyses set the groundwork for the future of this measurement with more capable detectors, such as the IceCube Upgrade and the proposed PINGU detector.
\end{abstract}

\section{Introduction}
\label{sec:intro}

The question of the Neutrino Mass Ordering (NMO) is one of the main drivers of the field of neutrino oscillation physics. The NMO describes the ordering of the three neutrino mass eigenstates $m_1$, $m_2$, and $m_3$. The two possible scenarios depend on the sign of $\Delta m_{31}^2 = m_3^2 - m_1^2$, often referred to as the atmospheric \textit{mass splitting}, 
where negative values are known as \textit{Inverted Ordering} (IO) and positive values as \textit{Normal Ordering} (NO).

The three neutrino mass states do not correspond directly to the three neutrino flavor states \nue, \numu, and \nutau. Instead, each mass state is a superposition of the flavour states, with the mixing described by the Pontecorvo-Maki-Nakagawa-Sakata (PMNS) matrix $U$~\cite{ref:Pontecorvo1,ref:Pontecorvo2,ref:MNS}, such that
\begin{align}
\nu_\alpha = \sum_{i=1}^3 U_{\alpha,i} \nu_i,
\end{align}
where $\alpha \in \left\{ e,\mu,\tau \right\}$ labels the flavor states and $i \in \left\{ 1,2,3 \right\}$ labels the mass states. By convention, $\nu_{1}$ is the state containing the most electron flavor, and $\nu_{3}$ is the state containing the least.

The mixing matrix $U$ can be parameterized by a CP-violating phase $\delta_\mathrm{CP}$ and three mixing angles $\theta_{12}$, $\theta_{13}$, and $\theta_{23}$. In the case of Majorana neutrinos, two additional phases are included, which are of no relevance for this work.
Since $U$ is non-diagonal, flavor changes are observed depending on the energy and propagation distance of a neutrino, which are commonly known as neutrino oscillations. The oscillations are described by the mass splittings, mixing angles, and the CP-violating phase\,\cite{Yanez:2015uta}. 

For propagation through dense matter, the neutrino oscillations are modulated by interactions with electrons, which give rise to matter effects~\cite{MSW_original_paper_v2} such as the so-called \textit{MSW effect} and 
\textit{parametric enhancement}\,\cite{Mikheev:1986gs, Mikheyev1986, akhmedov1998parametric,Petcov:1998su,Chizhov:1999he}. 
Depending on the NMO, these modulations arise mainly in the neutrino (NO) or anti-neutrino channel (IO)\,\cite{mena2008neutrino}.  In measurements of solar neutrino oscillations, they were used to determine the ordering of the neutrino states $\nu_1$ and $\nu_2$ by finding $m_2 > m_1$.
Moreover, these modulations can be observed for atmospheric neutrinos that undergo matter effects during their propagation through the Earth. 
In contrast to long-baseline accelerator experiments, the signature observed in IceCube is largely independent of the value of $\delta_\mathrm{CP}$, which allows for a complementary measurement of the NMO at higher energies, using atmospheric neutrinos\,\cite{Qian:2015waa}.

\begin{figure}[tb]
  \begin{center}
\subfigure[Normal Ordering]{
    \includegraphics[width=0.49\textwidth]{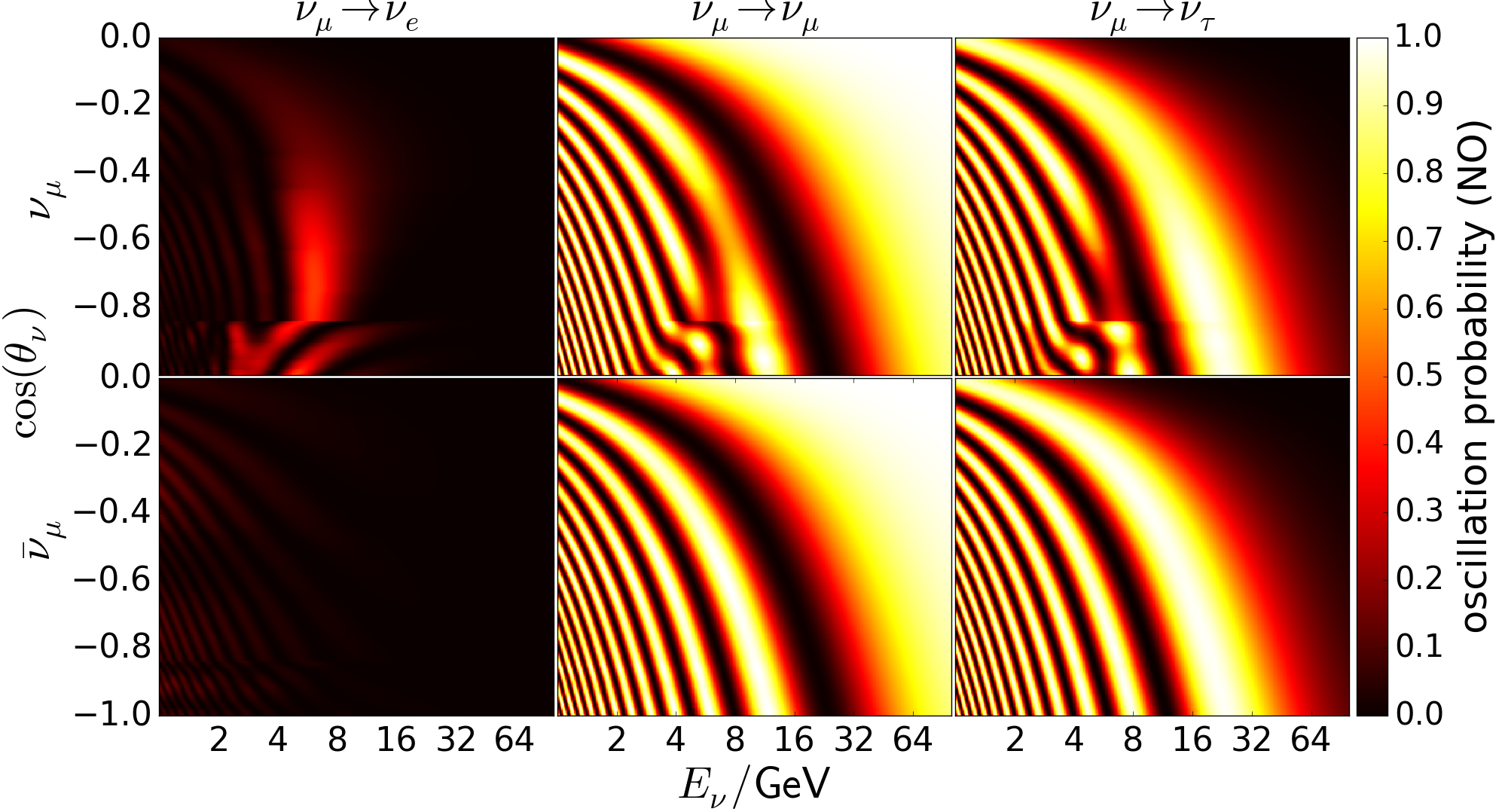} 
}
\subfigure[Inverted Ordering]{
    \includegraphics[width=0.49\textwidth]{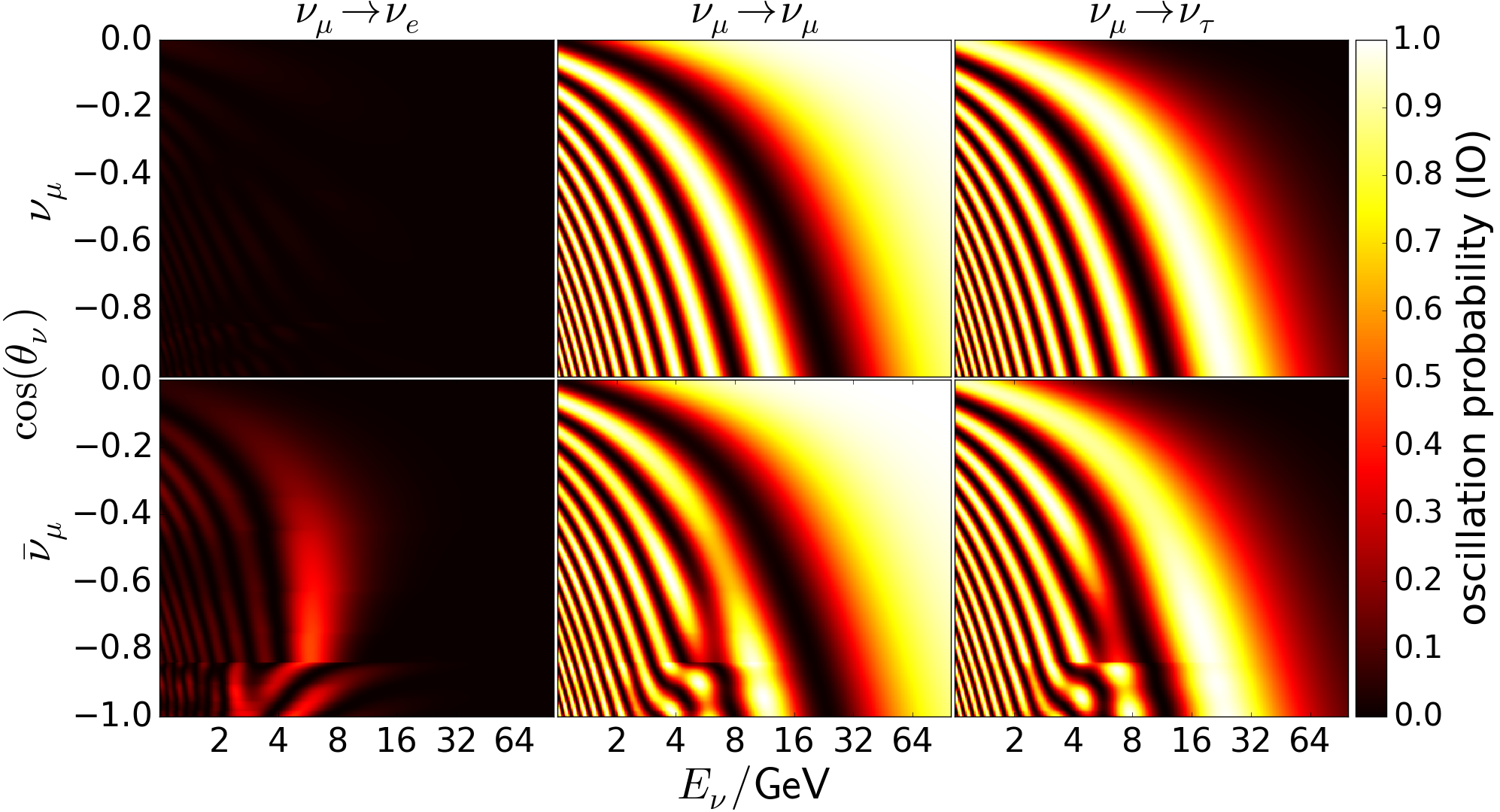} 
}
  \end{center}
  \caption{
 Oscillation probabilities for an atmospheric \numu or $\overline{\nu}_{\mu}$ upon reaching the IceCube detector, as a function of the cosine of the zenith angle, $\theta_\nu$, and the energy, $E_\nu$, of the neutrino, for the NO (a) and the IO (b) hypotheses. The probabilities are shown for the neutrino appearing as each of the three possible flavors, with the neutrino and anti-neutrino cases shown as the top and bottom rows in each panel. The dominant mixing of $\nu_\mu$ and $\nu_\tau$ is clearly visible, while the $\nu_e$ flavor is mostly decoupled, except for a small contribution from matter effects below $E_\nu \sim 15\,\mathrm{GeV}$.}
  \label{fig:Oscillograms}
\end{figure}

Atmospheric neutrinos are produced in the Earth's atmosphere by interactions of cosmic rays with the nucleons of the air, generating mesons. These mesons decay generating electron and muon (anti-)neutrinos, which propagate through the Earth and can eventually be detected by an underground neutrino detector, such as IceCube\,\cite{ IceCubeDetectorPaper}. 
The baseline of propagation through Earth can be inferred by measuring the incoming zenith angle of the neutrino.
The highest-energy oscillation maximum arises at 
$E_\nu\sim 25\,\mathrm{GeV}$ for vertically up-going neutrinos, moving to lower energy at shorter baselines towards the horizon. 
For energies above a few GeV, the oscillations are mostly driven by the parameters  $\theta_{23}$ and $\Delta m_{31}^2$, which are therefore referred to as \textit{atmospheric oscillation parameters}\,\cite{Yanez:2015uta}, while for vacuum only oscillations the value of $\theta_{13}$ is too small for any detectable effect. Considering matter effects,
however, the effective value of $\theta_{13}$  under the right conditions can become sizeable, resulting in oscillation with electron flavors as shown in Fig. \ref{fig:Oscillograms}.

\begin{figure}[tb]
  \begin{center}
    \includegraphics[width=0.48\textwidth]{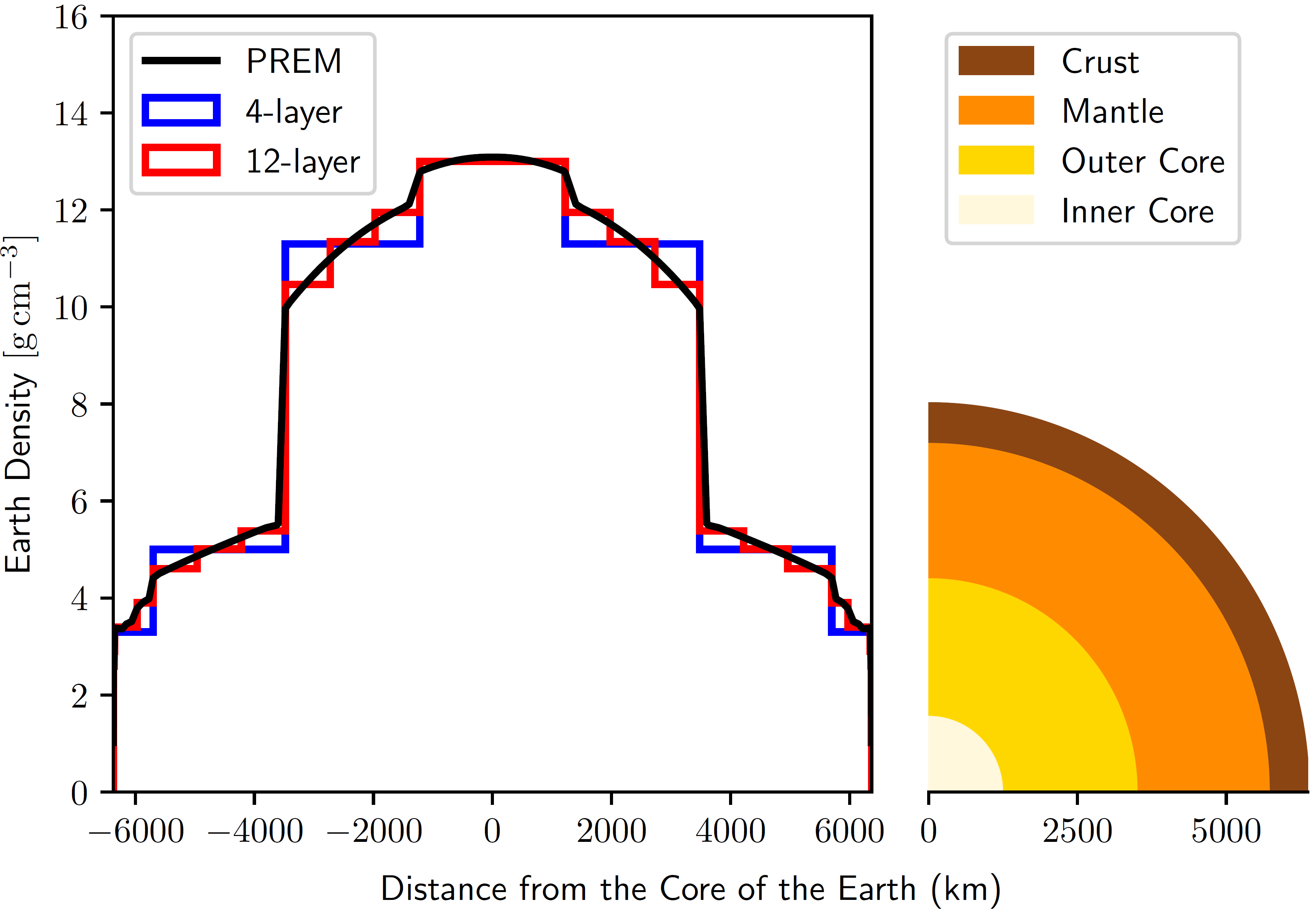} 
  \end{center}
  \caption{Earth density profile, according to the \textit{Preliminary Reference Earth Model} (PREM) and its approximation by 4- and 12-layers of constant density (commonly called PREM4 and PREM12, respectively)~\protect\cite{ref:PREM}.  }
  \label{fig:PREM}
\end{figure}

In atmospheric oscillations, the impact of the presence of matter arises mainly below $E_\nu\sim 15\,\mathrm{GeV}$. The strength of these matter effects depends on the Earth's matter profile, which we take as given by the Preliminary Reference Earth Model (PREM), shown in Figure~\ref{fig:PREM}~\cite{ref:PREM}. 

The oscillation probabilities for muon-flavored atmospheric neutrinos and anti-neutrinos 
to be found in the flavor state $\alpha \in \left\{ e, \mu, \tau \right\}$ 
for a given zenith angle $\theta_\nu$, and neutrino energy $E_\nu$, are shown in Figure~\ref{fig:Oscillograms}.
They are calculated with the \textit{PROB3++}\,\cite{prob3} package and the PREM12 approximation (cf. Figure~\ref{fig:PREM}), which are consistently used throughout this work.
Due to the Earth's geometry and its core-mantle structure, the visible modulations of atmospheric neutrino oscillations feature a clear zenith-dependence.

Note that the oscillation patterns for neutrinos and anti-neutrinos flip between the two orderings. Thus, the NMO can be determined by finding the enhancement in transition probabilities from matter effects either in the neutrino channel (NO) or anti-neutrino channel (IO).
For detectors insensitive to distinguishing 
neutrinos from anti-neutrinos on an event-by-event level, the NMO still leads to a visible net-effect in the amplitude of the observed matter effects, because the atmospheric fluxes and the cross sections for neutrinos and anti-neutrinos differ\,\cite{ref:PINGULoI, akhmedov2013mass}. 
These differences mean that
atmospheric neutrinos are measured at higher rates than the corresponding anti-neutrinos. Due to this rate difference, the strength of observed matter effects in a combined sample of neutrinos and anti-neutrinos is increased in case of NO and decreased in case of IO, which is the main signature targeted in this work.


The determination of the NMO has important implications for searches for neutrinoless double-$\beta$ decay, where the entire mass region allowed in the case of IO is in reach of the next generation of experiments\,\cite{ref:PDG, giuliani_andrea_2018}. 
The NMO must also be determined as part of the search for CP-violation in the lepton sector, where the sensitivity to $\delta_\mathrm{CP}$ depends strongly on the ordering\,\cite{Hagedorn:2017wjy, branco2012leptonic}. 
Therefore, a measurement of the NMO is targeted by several future long-baseline, reactor, and atmospheric neutrino experiments, such as DUNE\,\cite{DUNESensitivity}, JUNO\,\cite{JUNO_all},  PINGU\,\cite{ref:PINGULoI,ref:PINGUVision}, ORCA~\cite{ref:ORCA}, and Hyper-Kamio\-kande~\cite{ref:HyperKKorea}.
Moreover, current neutrino experiments such as T2K\,\cite{T2K_newest}, NOvA\,\cite{NOVA_newest},
 and Super-Kamiokande\,\cite{SuperK_newest} provide first indications of the NMO. 
Combining the results from several experiments, recent global fits prefer Normal over Inverted Ordering  at $\sim\!\! 2 - 3.5\,\sigma$ with a small preference for the upper octant (i.e.  $\sin^2(\theta_{23})>0.5$)\,\cite{ref:Lisi,NuFit_Web, NuFit_Paper, deSalas:2017kay}.


\section{The IceCube Neutrino Observatory}
\label{sec:icecube}

\begin{sloppypar}
The \textit{IceCube Neutrino Observatory}\,\cite{IceCubeDetectorPaper} is a $\sim\!\! 1\,\mathrm{km}^3$ neutrino detector at the Geographic South Pole, 
optimized for 
detecting atmospheric and astrophysical neutrinos above \mbox{$E_\nu\sim 100 \,\mathrm{GeV}$}. It consists of 86 strings running through the ice vertically from the surface almost to the bedrock, carrying a total of 5160 Digital Optical Modules (DOMs) at depths between $1450\,\mathrm{m}$ and $2450\,\mathrm{m}$ \cite{Aartsen:2013rt}. Each DOM houses a 10'' photomultiplier tube and digitizing electronics, surrounded by a glass sphere\,\cite{IceCubeDetectorPaper,Abbasi:2010vc,Abbasi:2008aa}. 
\end{sloppypar}


In the center of the detector, some of these strings form a more densely instrumented volume called \textit{DeepCore}\,\cite{ref:DeepCoreDetector}. It consists of 8 strings with an increased vertical density of DOMs with higher quantum-efficiency, surrounding one IceCube string. Due to the denser instrumentation and the higher quantum-efficiency DOMs, the DeepCore infill has a lower energy threshold than the surrounding IceCube array. The corresponding detection efficiency of DeepCore increases steeply between $\sim3\,\mathrm{GeV}$ and $\sim 10\,\mathrm{GeV}$ and flattens for higher energies\,\cite{IceCubeDetectorPaper, ref:DeepCoreDetector}.



Neutrinos are detected by the Cherenkov emissions of their charged secondary particles, which are generated by Charged Current (CC) and Neutral Current (NC) interactions with the nucleons of the ice. In the case of CC muon-neutrino interactions, a hadronic cascade is initiated  at the primary vertex, combined with an outgoing muon. The muon can propagate large distances through the detector, leading to an elongated shape of the energy deposition and thus of the Cherenkov light emission.  Such events are called \textit{track-like} signatures. In contrast, CC electron-neutrino, NC, and the majority of CC tau-neutrino interactions, do not produce a muon that can travel large distances. Instead, they initiate an electromagnetic and/or hadronic cascade that develops over a distance of a few meters. The light emission of this cascade is considerably smeared around the Cherenkov angle of the shower direction. Such events are called \textit{cascade-like}. 
At low energies below a few tens of GeV, the separation of track- and cascade-like events becomes increasingly difficult, due to the short muon track and the coarse detector granularity.
For oscillation measurements with DeepCore, this separation of track-like and cascade-like events is used to partially distinguish neutrino flavors\,\cite{ref:DeepCoreDetector}.


\newtext{For the analyses presented here, we use the Honda 3D atmospheric neutrino simulation~\cite{HONDA_2015}, and the GENIE neutrino interaction generator~\cite{axialMass} with KNO~\cite{ref:KNO} and PYTHIA~\cite{ref:Pythia}. For quasielastic and resonance events, the axial masses are set to \mbox{$M_{A}^\mathrm{qe}=0.99\,\mathrm{GeV}$} and \mbox{$M_{A}^\mathrm{res}=1.12\,\mathrm{GeV}$}, respectively. Simulation of the atmospheric muon background uses CORSIKA~\cite{ref:Corsika}, with the Polygonato-H\"orandel model of the muon energy spectrum~\cite{ref:PolygonatoHoerandel}. Muons are propagated through the ice using PROPOSAL~\cite{ref:Proposal}; the propagation of all other particles is based on GEANT4~\cite{ref:Geant4,ref:EMParameterization}. Cherenkov photons are propagated throught the ice using a GPU-based code~\cite{ref:GPUPhotons}. More details of the simulation can be found in~\cite{TauApp}.}

\section{Data Samples and Reconstruction} \label{sec:SampleReco}

In this work, two independent likelihood analyses are used to extract information about the NMO from DeepCore data. They are henceforth labelled Analysis~\greco and~\dragon, and the main differences between the two analyses are summarized in Table~\ref{tab:differences}. 
Analysis~\greco\ is designed to optimize the sensitivity to the NMO with DeepCore and considered the main result of this work, while Analysis~\dragon\ is designed to resemble the proposed PINGU analysis from~\cite{ref:PINGUVision}, \newtext{using only events that are fully-contained in the DeepCore detector}, and is used as a confirmatory result here. Further details about Analyses~\greco and~\dragon can be found in~\cite{LeuermannThesis} and~\cite{WrenThesis}, respectively. \newtext{The use of two independent analyses with  partially complementary data sets gives great confidence in the quantitative conclusions of the analysis presented here and the treatment and impact of the systematic uncertainties.}

The analyses are based on DeepCore data taken between May 2012 and April 2014, comprising a total livetime of $1006$ ($1022$) days for Analysis~\greco\ (\dragon). 
The difference in livetime arises from slightly different criteria on the stability of data acquisition. The data is run through two largely independent processing chains, where  
both samples are acquired by filtering the data in several successive steps of selection. These steps include the application of selection criteria on well-understood variables, as well as machine-learning methods, namely Boosted Decision Trees\,\cite{freund1997decision}. The selections are aiming for a reduction of the background of atmospheric muons and triggered noise, while maintaining a large fraction of well-reconstructed, low-energy neutrino events below \mbox{$E_{\nu}\sim 100\,\mathrm{GeV}$}.
Both samples are described in more detail in~\cite{TauApp}. 
Compared to~\cite{TauApp}, the samples used in this work differ by the following modifications:

\begin{table*}[htbp]
\begin{center}
\caption{ 
Overview of the main differences between the two NMO analyses in terms of 
the total number of observed events,
the selection strategy, the reconstruction likelihood, the reconstructed energy range, the number of analysis bins (given as number of $E_\nu^\mathrm{reco}, \vartheta_\nu^\mathrm{reco}, \mathrm{PID}$ bins), the background (atmospheric muon) description, the template generation, and the estimated fractions of the data sample from each contribution.}
\label{tab:differences} 
\centering
\begin{tabular}{ccc cc cc cc }
\bottomrule
 & Data  & Selection & Recon. & Energy  & Analysis & Background & Template & Estimated Contributions [\%]  \\
 & Events & Strategy & Likelihood &  Range & Binning & Description & Generation & $\mathrm{CC}\nu_e/\mathrm{CC}\nu_\mu/\mathrm{CC}\nu_\tau/\mathrm{NC}/\mu/\mathrm{noise}$   \\ 
\midrule
\greco
& $ 43\,214$  & high statistics & hit-based& $4-90\,\mathrm{GeV}$ & 10, 10, 3 & simulation & KDEs & 21.7 / 58.4 / 6.2 / ~~8.8 / 4.8 / 0.1  \\
\dragon 
& $23\,053$ & quality events & charge-based & $5-80\,\mathrm{GeV}$ & 10, ~~5, 2 & data & histograms & 29.4 / 58.0 / 2.0 / 10.4 / 0.2 / ~--~  \\
\bottomrule
\end{tabular}
\end{center}
\end{table*} 
 
First, 
\newtext{events with a reconstructed vertex outside the detector that enter from below}
are not vetoed in Analysis~\greco\ using the lower part of the DeepCore detector, as it is done for downgoing and horizontal events using the surrounding IceCube detector. This increases the statistics  at the expense of a reduced energy resolution for these uncontained events, especially at high energies. The loss in energy resolution is due to the unobserved fraction of deposited energy outside the detector volume. 
Second, the range of reconstructed energies considered is extended for both analyses compared to~\cite{TauApp}, from $56\,\mathrm{GeV}$ to $90\,\mathrm{GeV}$ ($80\,\mathrm{GeV}$) for Analysis~\greco\ (\dragon), allowing us to constrain nuisance parameters outside the strongest oscillation region. Third, both analyses use exclusively upgoing events (i.e.  $\cos(\theta_\nu^\mathrm{reco}) < 0$) to reduce the background from atmospheric muons. 
 
\begin{figure}
    \centering
    \includegraphics[width=\columnwidth]{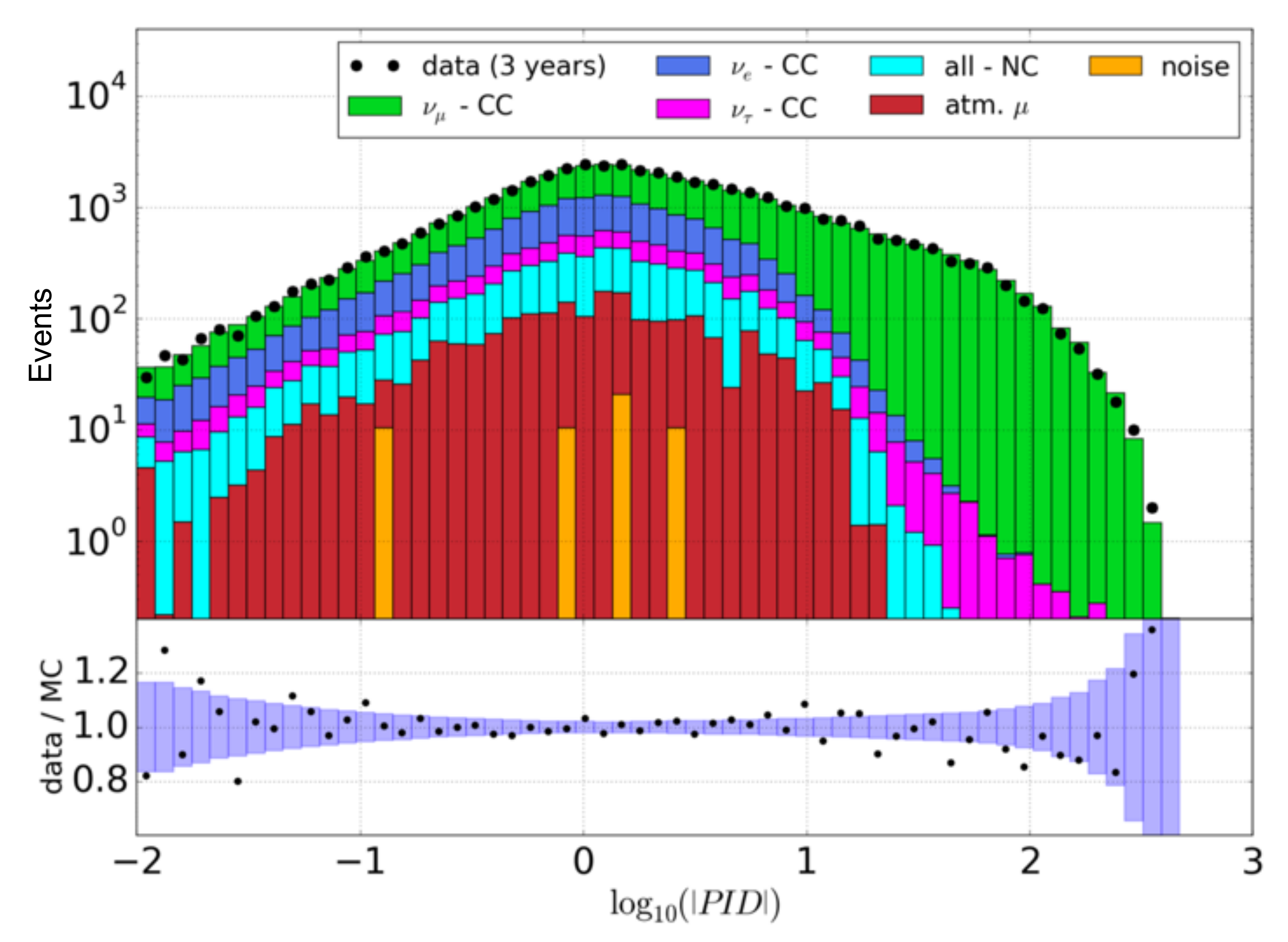}
    \caption{\newtext{The distribution of the particle identification variable for Analysis~\greco. The blue band on the data/MC ratio is the statistical uncertainty.}}
    \label{fig:PID}
\end{figure}

The final samples are reconstructed with the same algorithm for Analyses~\greco\ and~\dragon\,\cite{TauApp, LeuermannThesis}. 
It is based on a likelihood function that links the number and the arrival times of the observed Cherenkov photons in all DOMs to a physics hypothesis. The physics hypothesis is given by the position and time of the interaction vertex, the neutrino direction, and the neutrino energy, which are the parameters of the likelihood optimization. 
The reconstruction is run separately for a starting track  and a cascade-only hypothesis, where the starting track hypothesis features a cascade at the primary vertex with an additional parameter $L$ for the length of an outgoing muon track. Since the track hypothesis allows for fitting the track length to $L=0$, the 7-dimensional cascade-only-hypothesis is nested within the 8-dimensional track-hypothesis. The log-likelihood difference between track and cascade-only hypothesis is used as the flavor-separating variable, called \textit{Particle Identification} (PID). Besides the reconstructed neutrino zenith angle $\theta_\nu^\mathrm{reco}$ and neutrino energy $E_\nu^\mathrm{reco}$, the PID is used as a third observable entering the likelihood analyses described in Section~\ref{sec:Analyses}.
\newtext{The distribution of the PID variable for Analysis~\greco is shown in Fig.~\ref{fig:PID}.}

In the reconstruction, the optimized likelihood function differs between the two analyses:
For Analysis~\dragon, the reconstruction likelihood is defined using the observed charge binned in time for each DOM as a proxy for the observed number of Cherenkov photons. Since some deviations were found between data and Monte Carlo in charge-related quantities, the likelihood was reformulated in a charge-independent way for Analysis~\greco, such that the charge amplitude information was removed and the only information used is whether a DOM is hit or not hit in multiple bins of time.
In terms of the resolutions in reconstructed zenith angle $\theta_\nu^\mathrm{reco}$ and neutrino energy $E_\nu$, the impact of the likelihood reformulation was found to be small. Moreover for Analysis~\dragon, the impact of the charge mismatch is estimated to be small in comparison to the statistical uncertainty on the observed NMO.

After the data selection, the number of events in Sample~\greco\ exceed the number of events in Sample~\dragon\ by a factor of $1.87$, while providing similar resolutions in energy and zenith angle.

Note that for Analysis~\dragon, the atmospheric muon background is estimated from data in an off-signal region, while for Analysis~\greco, it is obtained from Monte Carlo simulations (cf. Table~\ref{tab:differences}). As a result, there is no \textit{a priori} Monte Carlo prediction for the atmospheric muon contamination in Sample~\dragon. However, the fraction of atmospheric muons is fitted in the analysis as discussed in Section~\ref{sec:firstresults}.
The contamination of triggered noise was found to be only $\lesssim 0.1\%$ for both samples. It is included into the likelihood fit for Analysis~\greco, while it is neglected for Analysis~\dragon. 

The final samples consist of CC muon neutrino, CC electron neutrino, CC tau neutrino, NC, and atmospheric muon events.
These different components are called \textit{contributions} in the following and are simulated separately in Monte Carlo except for the atmospheric muon contribution used in Analysis~\dragon
that is parametrized from an off-signal data region.

The estimated fraction of the data samples from each contribution is shown in Table~\ref{tab:differences}. These fractions are calculated using the best-fit values for all systematic parameters, discussed in Section~\ref{sec:Analyses}.

\section{Analyses} \label{sec:Analyses}

\noindent  Both Analyses~\greco\ and~\dragon\ use a binned likelihood method to determine the NMO by observing the signature from Figure~\ref{fig:Oscillograms} in reconstructed variables. 
Since a separation of all flavors cannot be done with DeepCore, the PID is used to distinguish track- and cascade-like events, while neutrino energy and zenith angle are obtained from the reconstruction described in Section~\ref{sec:SampleReco}.

\begin{figure}[tb]
  \begin{center}
    \includegraphics[width=0.49\textwidth]{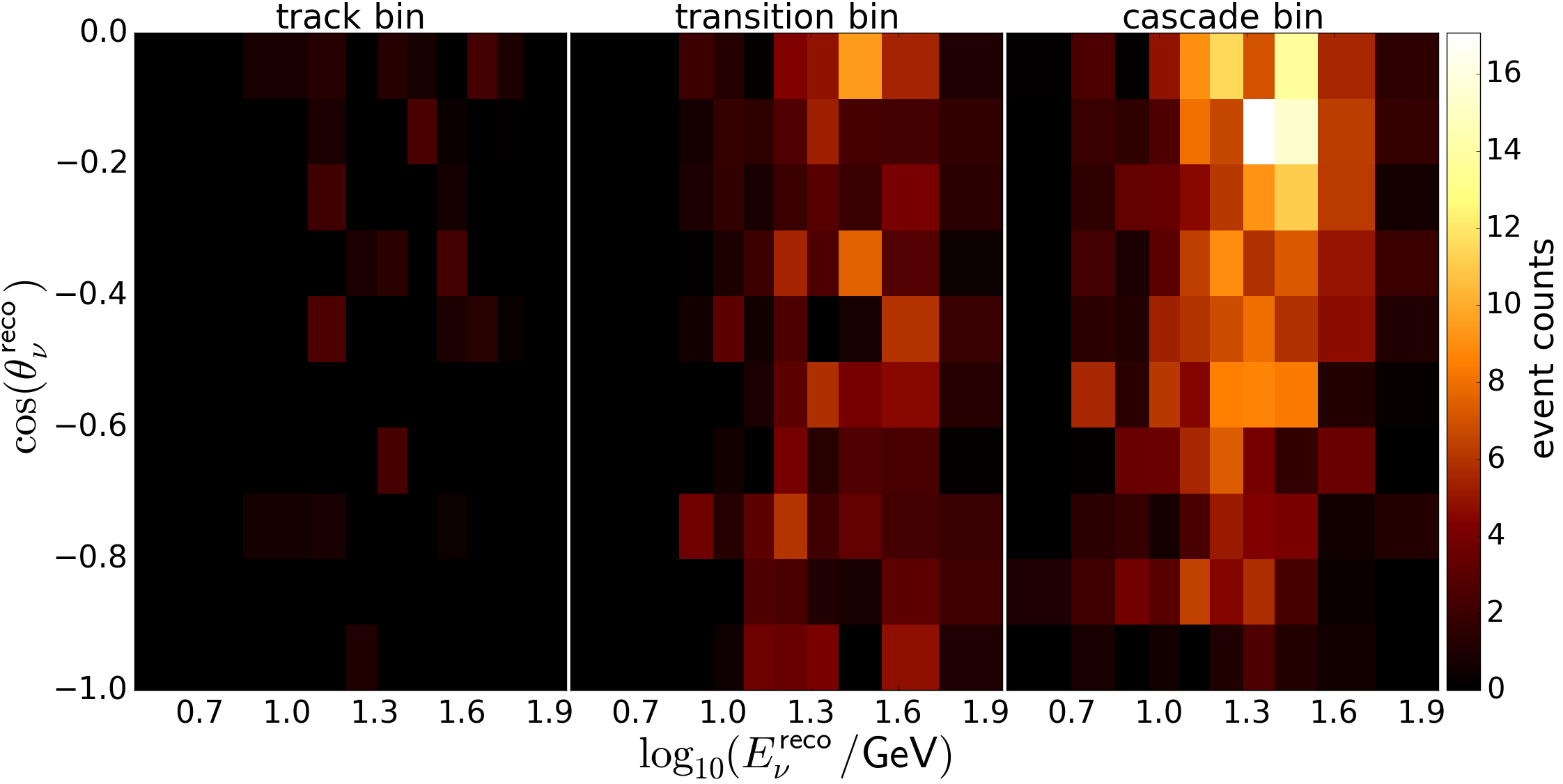} 
    \includegraphics[width=0.49\textwidth]{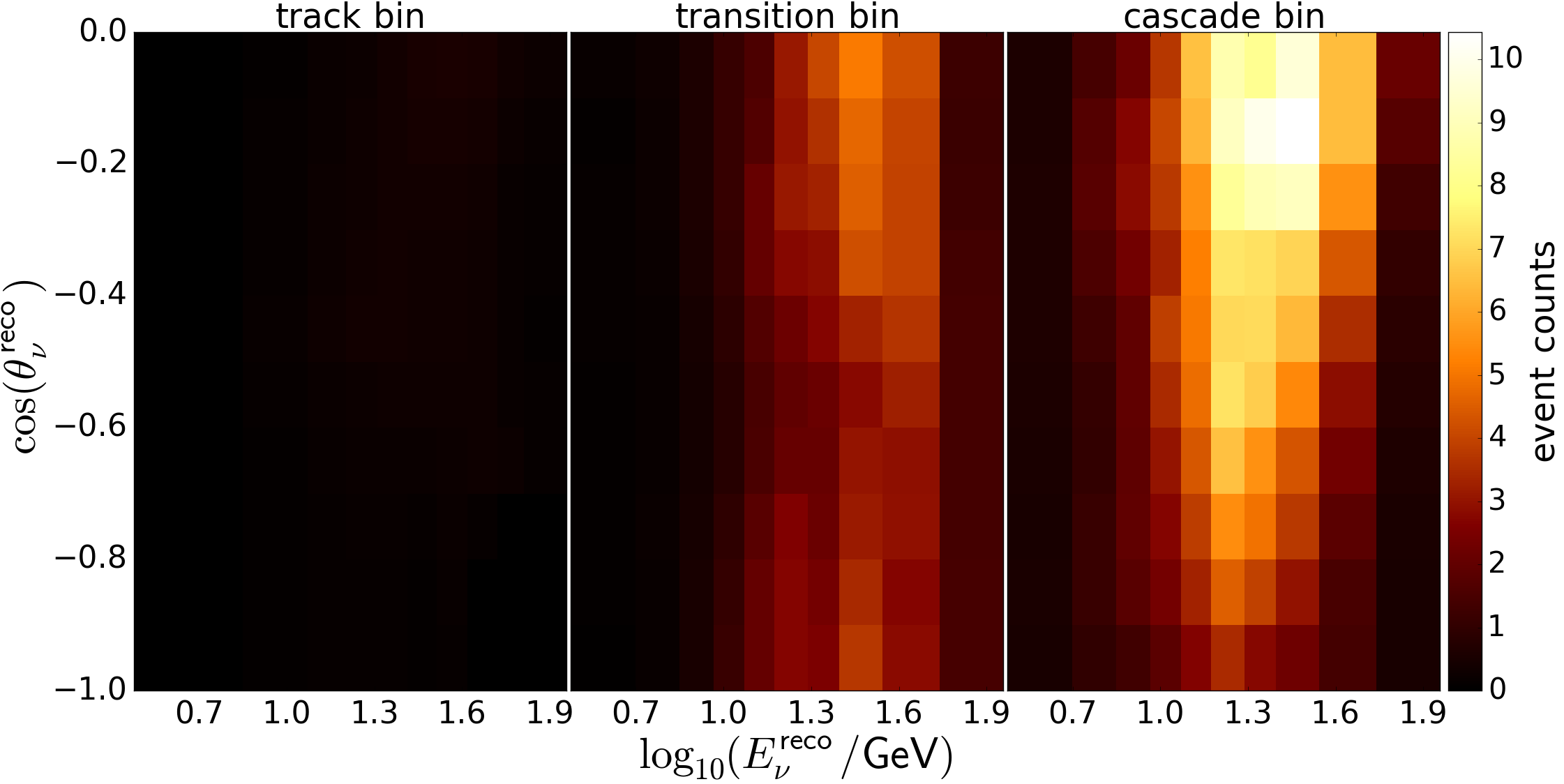} 
  \end{center}
  \caption{Comparison of Monte Carlo template for atmospheric muons in  Analysis~\greco, generated as histogram (top) and KDE (bottom): the latter one is used in the analysis, due to the reduced impact of limited Monte Carlo statistics. }
  \label{fig:KDETemplate}
\end{figure}

For both analyses, the binning is summarized in Table~\ref{tab:differences}. For Analysis~\dragon\ only two PID bins are used to separate track- and cascade-like events, analogously to~\cite{ref:PINGULoI}, while Analysis~\greco\ uses three PID bins. This is motivated by the weak separation power at low energies, where the confidence in the separation can be taken into account by including an additional, intermediate PID bin.
The binning in neutrino energy and zenith angle is chosen to be uniform in $\log_{10}(E_\nu^\mathrm{reco})$ and $\cos(\theta^\mathrm{reco}_\nu)$ for Analysis~\dragon. For Analysis~\greco, it is also uniform in $\cos(\theta^\mathrm{reco}_\nu)$, while it is optimized in $\log_{10}(E_\nu^\mathrm{reco})$ to roughly follow the available statistics and maintain a large number of bins in the most interesting region at $E_\nu \sim 10\,\mathrm{GeV}$.

In Analysis~\dragon, the binning is used to generate Monte Carlo distributions, called \textit{templates}, in $E_{\nu}^\mathrm{reco}$, $\theta_{\nu}^\mathrm{reco}$, and PID for each contribution to the data sample, using histograms. In contrast, Analysis~\greco\ applies an adaptive \textit{Kernel Density Estimation} (KDE) method to produce these templates, which smooths the 
fluctuations 
from limited Monte Carlo statistics. These uncertainties arise mainly from the atmospheric muon template, where the available Monte Carlo statistics are similar to those from experimental data, due to the time-intensive simulation of atmospheric muons. 

The KDE method is analogous to the one used in~\cite{DiffuseFlux} and based on~\cite{KDE_thisMethod}. However, the method from~\cite{KDE_thisMethod} is extended by reflecting the KDE at the boundaries of the binned parameter space and integrating the resulting distribution to obtain a prediction for the bin content\,\cite{kde_reflect}.
For the atmospheric muons, this is illustrated in Figure~\ref{fig:KDETemplate}, where the Monte Carlo template for atmospheric muons is generated with histograms (top) and the above mentioned KDE method (bottom). In the case of histograms, the fluctuations in the bin content, arising from limited Monte Carlo statistics, are clearly visible. 

The uncertainties on the KDE prediction are estimated using \textit{bootstrapping} for every contribution from Section~\ref{sec:SampleReco} separately\,\cite{bootstrap}.
For each contribution, which consists of $N$ MC events, events are drawn randomly from this sample, replacing the event each time so that it can be drawn again, until $N$ events have been drawn. This new sample of $N$ events is called a bootstrapped sub-sample, and from this a new KDE template is generated. This process is repeated several times and the uncertainty on each bin content in the original KDE template is estimated from the resulting distribution of bin contents in the bootstrapped samples.

\begin{figure}[tb]
  \begin{center}
    \includegraphics[width=0.5\textwidth]{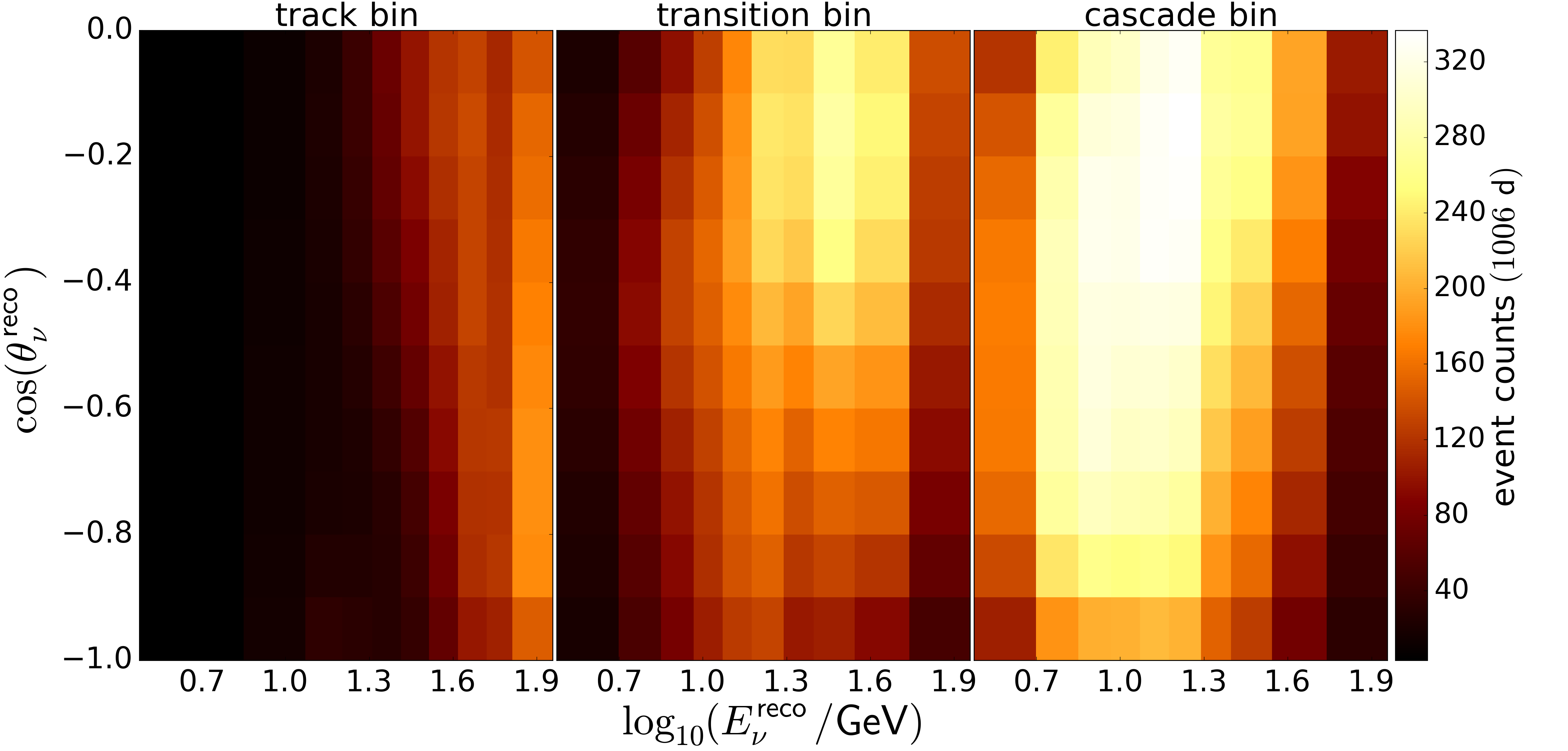} 
    \includegraphics[width=0.5\textwidth]{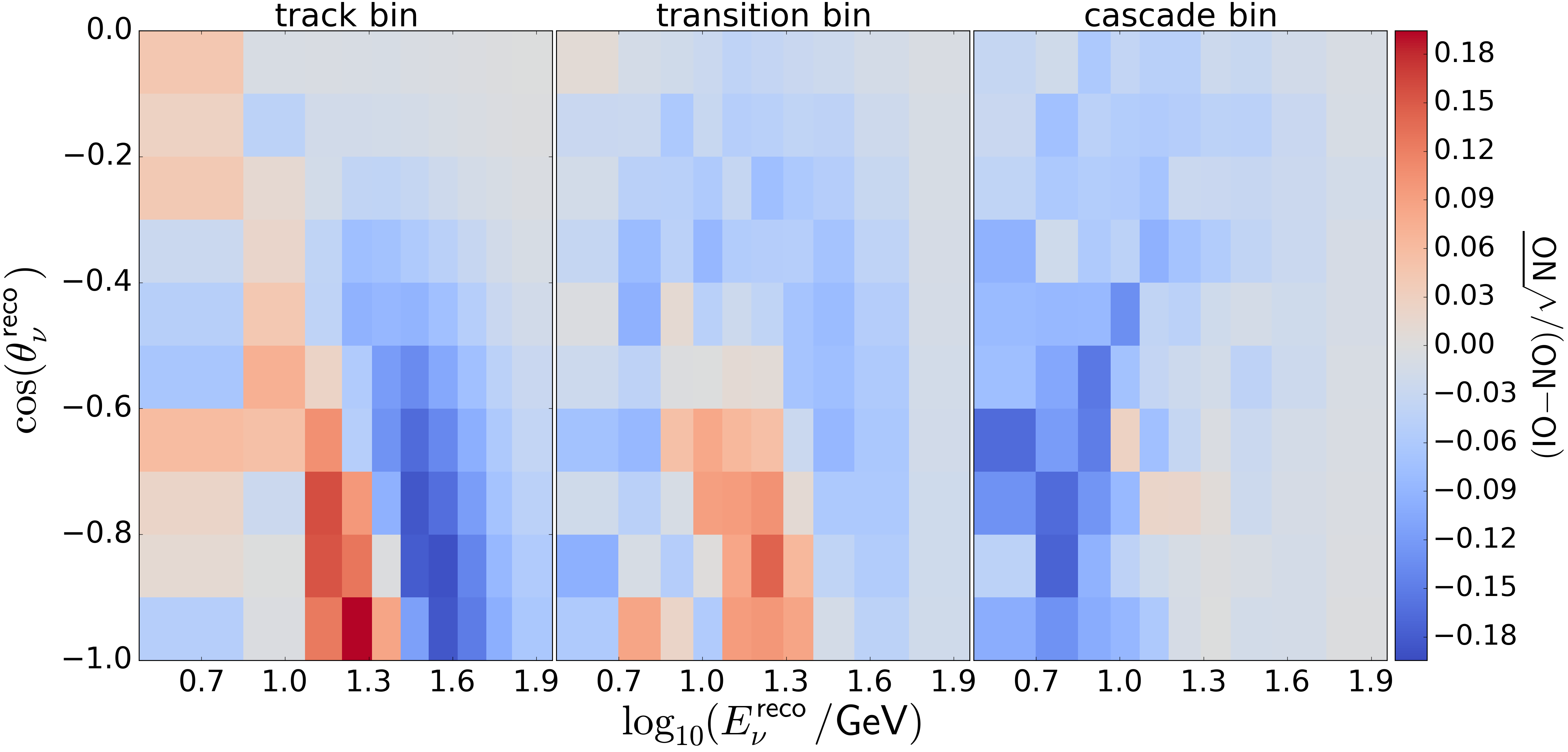} 
  \end{center}
  \caption{Top: the distribution in PID, zenith angle, and neutrino energy for 
  Analysis~\greco 
  that enters the likelihood calculation; bottom: corresponding signature of the NMO, given as expected pull on the bin content in case IO is observed but NO is tested, using Poissonian statistics.
  }
  \label{fig:GRECOTemplates}
\end{figure}

For Analysis~\greco, the three-dimensional template obtained from the combination of all Monte Carlo contributions is shown in Figure~\ref{fig:GRECOTemplates}. 
Additionally, the expected pulls on each bin are shown in the case that the true ordering is inverted but the NO hypothesis is tested. This is used as a metric to visualize the signature of the NMO\,\cite{ref:PINGULoI}. 
As can be seen in Figure~\ref{fig:GRECOTemplates}, the expected pulls between NO and IO are small, which already indicates the low sensitivity due to the limited resolution and statistics of DeepCore at energies $E_\nu \lesssim 15\,\mathrm{GeV}$.

Using these distributions, likelihoods are defined for both analyses. 
For Analysis~\greco, the negative log-likelihood LLH is given by
\begin{equation}  \label{eq:LLHGRECO}
\mathrm{LLH} =  \left[ - \sum_{i \in \left\{ \mathrm{bins} \right\} } 
\ln\left( \frac{p_i^\mathrm{tot}(N^\greco_i, \mu^\greco_i, \sigma^\greco_{\mu_i})}{p_i^\mathrm{tot}(N^\greco_i, N^\greco_i, \sigma^\greco_{\mu_i})}  \right) \right] 
+ \frac{1}{2} S,
\end{equation}
where the term $S$ is common to the likelihood of both analyses and will be defined after discussing the other terms. The term $p_i^\mathrm{tot}(N^\greco_i, \mu^\greco_i, \sigma^\greco_{\mu_i})$ gives the probability of observing $N^\greco_i$ events in bin $i$, if $\mu^\greco_i$ events are expected. It is obtained by a convolution of a Poissonian distribution and a narrow log-normal probability density function that describes the uncertainty $\sigma^\greco_{\mu_i}$ on the Monte Carlo prediction $\mu_i$.
The uncertainty $\sigma_{\mu_i}^\greco$ is obtained from a quadratic combination of the individual template uncertainties for every contribution, obtained from bootstraping. 

Due to the KDE method used in Analysis~\greco, the dominant template uncertainties in the description of atmospheric muons are strongly reduced, such that the uncertainties on the total template are typically $\sim 10\%$ of the Poissonian error expected from data fluctuations. Thus, for Analysis~\greco\ these template uncertainties contribute only marginally to the following results.

For Analysis~\dragon, the likelihood is adapted from~\cite{DRAGON_numu}, where a $\chi^2$-value is calculated by quadratically combining the Poissonian error on the predicted bin content $ \mu^{\dragon}_i $ with the uncertainty $\sigma^{\dragon}_{\mu_i}$ on the combined template of all contributions. It is given by
\begin{equation}  \label{eq:Chi2DRAGON}
\chi^2 =   2\mathrm{LLH} = \sum_{i \in \left\{ \mathrm{bins} \right\} } 
 \frac{\left(N^\dragon_i - \mu^\dragon_i\right)^2}{ \mu^\dragon_i +(\sigma^\dragon_{\mu_i})^2 }
+  S,
\end{equation}
where the labels are analogous to Equation~\ref{eq:LLHGRECO}. Here, the uncertainties $\sigma^{\dragon}_{\mu_i}$ on the templates are estimated from the statistical error due to limited Monte Carlo and an uncertainty on the atmospheric muon template, estimated from off-signal data.

\begin{table*}[ht]
\begin{center}
\begin{threeparttable}
\caption{\label{tab:systematics} Systematics treated as nuisance parameters in the likelihood analysis, including normalization (N), detector response (D), oscillation (O), flux (F), and neutrino-nucleon interaction (I) uncertainties. These parameters are discussed in more detail in~\cite{TauApp}. The table gives the baseline value and, if the parameter is used with a prior in the likelihood, the standard deviation of the Gaussian prior, as well as the experimental best-fit values for both analyses and ordering hypotheses.}
\label{tab:Params}
\centering
\begin{tabular}{lll ll cccc }
\bottomrule
Label & Type &  Description of Parameter & Baseline$\pm$Prior & \multicolumn{2}{c}{Analysis~\greco}  & \multicolumn{2}{c}{Analysis~\dragon}  \\ 
 & & & & NO & IO & NO & IO  \\
\midrule
$N_\nu$ & N, F & normalization of total neutrino template & $1$\tnotex{tnote:fon} \,\tnotex{tnote:free} & 0.83 & 0.84 & 0.98 & 0.99  \\
$N_{\nu_e}$ & N, F & normalization of $\nu_e$ flux before oscillations & $1 \pm 0.05$\tnotex{tnote:fon} \,\tnotex{tnote:noprior}  & 1.00 & 1.00 & 1.37 & 1.38 \\
$N_\mathrm{NC}$ & N, I & normalization of NC events & $1 \pm 0.2$\tnotex{tnote:fon} & 0.74 & 0.75 & 0.99 & 0.99 \\
$N_\mu$ & N, F & normalization of atmos. muon events & $1$\tnotex{tnote:fon} \,\tnotex{tnote:free}  & 1.35 & 1.34 & 0.2\%\tnotex{tnote:mus} & 0.2\%\tnotex{tnote:mus} \\
$\epsilon_\mathrm{opt}$ & D & overall optical efficiency\,\cite{IceCubeDetectorPaper} &  $1\pm 0.1$\tnotex{tnote:fon} \,\tnotex{tnote:noprior}  & 1.00 & 1.00 & 0.92 & 0.92 \\
$\epsilon_\mathrm{lateral}$ & D & lateral dependence of optical efficiency\,\cite{IceCubeDetectorPaper} &  $0\pm1\tnotex{tnote:au}$  & 0.68 & 0.68 & -0.46 & -0.46 \\
$\epsilon_\mathrm{head-on}$ & D & head-on optical efficiency\,\cite{IceCubeDetectorPaper} &  $0$\tnotex{tnote:au} \,\tnotex{tnote:free}   & -1.01 & -1.01 & -2.00 & -1.92 \\
$\Delta m_{31}^2 / (10^{-3}\,\mathrm{eV}^2)$  & O & atmospheric mass-splitting &  $2.5$(NO)$/-2.43$(IO)\tnotex{tnote:free}  & 2.626 & -2.511 & 2.462 & -2.348\\
$\sin^2(\theta_{23})$  & O & atmospheric neutrino mixing angle &  $0.455\tnotex{tnote:free}$ & 0.476 & 0.485 & 0.558  & 0.539 \\
$\gamma_{\nu}$  & F & neutrino spectral index unc.\,\cite{HONDA_2015} &  $0.0\pm 0.1$\tnotex{tnote:noprior} & 0.073 & 0.071 & -0.025 & -0.027  \\
$\gamma_{\mu}$  & F & atmospheric muon spectrum unc.\,\cite{WrenThesis,ref:CosmicPrimarySys} &  $ 0.0\pm1.0$\tnotex{tnote:au}  & 0.04 & 0.04 & -- & -- \\
$\sigma_\nu^\mathrm{zenith}$ & F & zenith-dependent unc. in $\nu/\bar{\nu}$ flux\,\cite{Barr}  &  $0.0\pm1.0$\tnotex{tnote:au} \,\tnotex{tnote:noprior} & -0.12 & -0.11 & -0.86 & -0.89 \\
$\Delta(\nu/\bar{\nu})$  & F & energy-dependent unc. in $\nu/\bar{\nu}$ ratio\,\cite{Barr} &  $0.0\pm1.0$\tnotex{tnote:au}& -1.03 & -1.02  & 0.05 & 0.07 \\
$M_A^\mathrm{res} / \mathrm{GeV}$& I & axial mass unc. of resonant events\,\cite{axialMass} &  $1.12\pm0.22$ & 1.091 & 1.095 & 1.003 & 0.999 \\
$M_A^\mathrm{qe} / \mathrm{GeV}$ & I & axial mass unc. of quasi-elastic events\,\cite{axialMass} &  $0.99\pm0.25$  & 0.862 & 0.867 & 0.881 & 0.888 \\
\bottomrule
\end{tabular}
 \begin{tablenotes}
      \item\label{tnote:fon} relative to the nominal value of this parameter
      \item\label{tnote:au} parametrized with respect to the value and the uncertainty obtained from the provided reference
      \item\label{tnote:mus} given as fraction of the total sample, since no Monte Carlo prediction exists to compare to
      \item\label{tnote:noprior} no prior used for likelihood in Analysis~\dragon
      \item\label{tnote:free} \newtext{parameter allowed to vary freely (no prior) in both analyses}
 \end{tablenotes}
\end{threeparttable}
\end{center}
\end{table*}

The dominant systematic uncertainties are included in both likelihood fits using nuisance parameters. 
These nuisance parameters comprise uncertainties in the atmospheric neutrino flux, the atmospheric oscillation parameters, the neutrino-nucleon cross sections, and the detector response. \newnewtext{All systematic parameters are allowed to vary simultaneously and independently in the fit; we assume there are no correlations between the pulls on the various  parameters.} The parameters are listed in Table~\ref{tab:systematics}. To account for external constraints on these systematic parameters, Gaussian priors are included into the likelihood by the term $S$,
\begin{equation}\label{eq:priorS}
S =\sum_{s \in \left\{ \mathrm{sys} \right\} } \left( \frac{ s-s_0}{\sigma_s} \right)^2,
\end{equation}
where the sum runs over all systematic parameters. For each parameter, the tested value $s$ is compared to the expected baseline value $s_0$ with respect to its estimated uncertainty $\sigma_s$.
The baseline value $s_0$ and width $\sigma_s$ of each prior are identical for both analyses, and are stated in Table~\ref{tab:systematics}; the central value and the width are motivated by the provided references where possible. As indicated in Table~\ref{tab:Params}, the prior for some parameters was removed in Analysis~\dragon.  Due to the small sensitivity to the NMO, the prior assumption was found to imply a preference on the NMO in case the true parameter value differs from the baseline value, which is avoided by removing the corresponding priors from the likelihood. Thus, no external knowledge is included on these parameters, allowing for larger deviations from the baseline value.

The parameters $N_\nu$, $N_{\nu_e}$, $N_\mathrm{NC}$, and $N_\mu$ are used to vary the normalizations of the different contributions from Table~\ref{tab:differences}. Thus, they account for uncertainties in interaction cross sections, the total neutrino and muon fluxes, the $\nu_e/\nu_\mu$ production ratio, and detection efficiencies. 

Additional uncertainties on the neutrino fluxes predicted in~\cite{HONDA_2015} are modelled by the parameters $\gamma_\nu$, $\sigma_\nu^\mathrm{zenith}$, and $\Delta(\nu/\bar{\nu})$. Here, $\gamma_\nu$ incorporates uncertainties in the neutrino energy spectrum, arising from flux, and cross section uncertainties, according to a reweighting of Monte Carlo events \mbox{$\propto\!(E_\nu/\mathrm{GeV})^{\gamma_\nu}$}, while $\sigma_\nu^\mathrm{zenith}$ and $\Delta(\nu/\bar{\nu})$ incorporate the dominant uncertainties from~\,\cite{Barr} in an \textit{ad hoc} parametrization. The uncertainties on the production of atmospheric muons arising from the spectrum and compositions of the cosmic ray primary flux are represented by the parameter $\gamma_\mu$. Note that $\gamma_\mu$ is only included as an uncertainty for Analysis~\greco, since the atmospheric muon template in Analysis~\dragon\ is estimated from data.

Uncertainties in neutrino-nucleon interactions are represented by the parameters $M_A^\mathrm{res}$ and $M_A^\mathrm{qe}$, which model the axial mass of resonant and quasi-elastic interactions. Note that uncertainties on the cross section for deep inelastic scattering were also parametrized, but found to be negligible and therefore not included into the likelihood fit.

Detector uncertainties are modelled by the parameters $\epsilon_\mathrm{opt}$, $\epsilon_\mathrm{lateral}$, and $\epsilon_\mathrm{head-on}$, which describe the optical detection efficiency of the DOMs.  The value of  $\epsilon_\mathrm{opt}$ gives the total detection efficiency per photon, relative to the baseline scenario. In contrast, the parameters $\epsilon_\mathrm{lateral}$ and $\epsilon_\mathrm{head-on}$ describe the dependence of the photon detection efficiency on the inclination angle of the incoming photon. Here, $\epsilon_\mathrm{lateral}$ changes the slope of the acceptance curve, while $\epsilon_\mathrm{head-on}$ controls the acceptance of very vertically upgoing photons independently. Besides actual uncertainties in the DOMs' detection efficiency, these parameters incorporate uncertainties with respect to the optical properties of the ice in the refrozen drill holes that surround the DOMs.


All of the systematic parameters mentioned above are described in more detail in~\cite{TauApp}.
\newtext{Besides the parameters included in the fit, additional uncertainties have been investigated and tested for their possible effect on the analysis \cite{LeuermannThesis,WrenThesis}.
These parameters are the normalizations of sub-dominant experimental backgrounds (detector noise and event pile-up from coincident atmospheric muons), additional uncertainties on the optical properties of the ice, the oscillation parameters  ($\theta_{12}$, $\theta_{13}$, $\Delta m_{21}^2$, and $\delta_\mathrm{CP}$), and Bjorken-$x$ dependent uncertainties in the cross section for deep-inelastic neutrino-nucleon scattering.}
\newnewtext{Two types of tests were performed to determine the impact of these parameters. In the first test, a parameter is injected into a MC fake dataset, shifted from its nominal value by $\pm 1\sigma$ in the case of a detector systematic and by $\pm 3\sigma$ in the case of an oscillation parameter. 
This MC fake data is fit using the same MC set, but with the parameter in question fixed to its unshifted nominal value to assess whether the uncertainty in the systematic parameter can bias the measured ordering hypothesis. This test is repeated for MC fake datasets generated with both mass orderings. For none of these systematic or oscillation parameters is a bias observed in the measured preference for the mass ordering of more than $0.05\sigma$.
For the case of $\delta_\mathrm{CP}$, the value of $\delta_\mathrm{CP}=180^{\circ}$ is being used as the `nominal' value for the MC fake dataset, with the value of $\delta_\mathrm{CP}=270^{\circ}$ injected into the MC used in the fit; a negligible (less than $0.01\sigma$) bias in the ordering preference is observed.
In the second test, the same shifted values as above are injected into the MC fake dataset, but now the parameter under test is allowed to vary in the fit. This allows us to determine if the inclusion of these parameters into the fit causes any loss of sensitivity to the mass ordering. None of the parameters in question cause a loss of sensitivity of more than $0.05\sigma$; the inclusion of $\delta_\mathrm{CP}$ reduces the sensitivity by less than $0.03\sigma$. Since all the parameters described in this paragraph are  shown to have no impact on the mass ordering sensitivity, or the potential to cause a bias, they have been set to their nominal values (in the case of oscillation parameters, to the NuFit~\cite{NuFit_Paper} best-fit values), and are not included in the final fit in order to minimise the computing time required for the multi-parameter minimisation.}


For Analysis~\greco\ (\dragon), the negative log-likelihood from Equation~\ref{eq:LLHGRECO}  (\ref{eq:Chi2DRAGON}) is optimized. To do this, $\mathrm{LLH} \equiv -0.5 \chi^2$ is used as the negative log-likelihood for Analysis~\dragon. 
 During this optimization, the first and the second octant in $\theta_{23}$ are fitted separately for both orderings, allowing all the parameters listed in Table~\ref{tab:systematics} to vary, and the fit optimizing the $\mathrm{LLH}$ is taken as the best-fit for this ordering. The resulting difference, $2\Delta \mathrm{LLH}_\mathrm{NO-IO} \equiv \Delta\chi^{2}_{\mathrm{NO}-\mathrm{IO}}$, between the NO and IO hypotheses is then calculated for both analyses.


Finally, $2\Delta \mathrm{LLH}_\mathrm{NO-IO}$  ($\chi^2_\mathrm{NO-IO}$) is used as a \textit{test-statistic} (TS) in Section~\ref{sec:firstresults} for Analyses~\greco\ (\dragon) to derive the experimental result from the fit to the data.

\section{Sensitivity to the Neutrino Mass Ordering} 
\label{sec:anatechnique}


\noindent 
The determination of the Neutrino Mass Ordering is a binary hypothesis test, which requires the test of two non-nested hypotheses. This is different from most other applications in particle physics, where a general hypothesis $\mathcal{H}_G$ is tested against a specific one, $\mathcal{H}_S$, in the sense that the specific hypothesis is obtained for a certain realization of the parameters of $\mathcal{H}_G$. For such nested hypotheses, \textit{Wilks' Theorem} is commonly used to derive sensitivities and to estimate limits on fitted parameters\,\cite{Wilks}.
In contrast, Wilks' Theorem does not apply to the determination of the Neutrino Mass Ordering, since the discrete choice of Normal or Inverted Ordering is not related to the fixing of degrees of freedom\,\cite{Ciuffoli:2013rza}.

Due to the subtleties involved in the statistical treatment and since a determination of the NMO is expected within the next decade, the correct method to quantify the preference is object of many discussions\,\cite{Blennow:2013oma, Qian:2012zn, Ciuffoli:2013rza}.
Here, two methods are used to estimate the sensitivity, which are described in the following.


The first method is a statistically rigorous analysis of the resulting likelihood values, using the obtained value of $2\Delta\mathrm{LLH}_\mathrm{NO-IO}$ as a TS. It derives the resulting sensitivity, given by the expected confidence in the determination of the NMO, from a frequentist coverage test. To do this, the data is fit with both ordering hypotheses giving a value for the TS and two sets of best-fit systematic parameters, $\eta^{\mathrm{NO}}$ and $\eta^{\mathrm{IO}}$. These fits are called \textit{fiducial fits} (FD) in the following.

From these parameters, the resulting best-fit templates are generated for NO and IO. Then, these templates are used to generate \textit{Pseudo-Experiments} or \textit{Pseudo-Trials} (PT)s by adding Poissonian fluctuations on the bin-contents, as expected in a real-world experiment; \newtext{in this analysis, which has a sensitivity dominated by the statistical uncertainty, it is unnecessary to fluctuate each PT according to the systematic uncertainties}. Afterwards, each PT is fitted with both ordering hypotheses, resulting in a new value for the $\mathrm{TS} = \Delta \chi^2_\mathrm{NO-IO}= 2\Delta\mathrm{LLH}_\mathrm{NO-IO}$. From these PTs, two distributions of the TS are obtained for the two sets of injected parameters $\eta^{\mathrm{NO}}$ and $\eta^{\mathrm{IO}}$.

\begin{figure}[tb]
  \begin{center}
    \includegraphics[height=5.5cm]{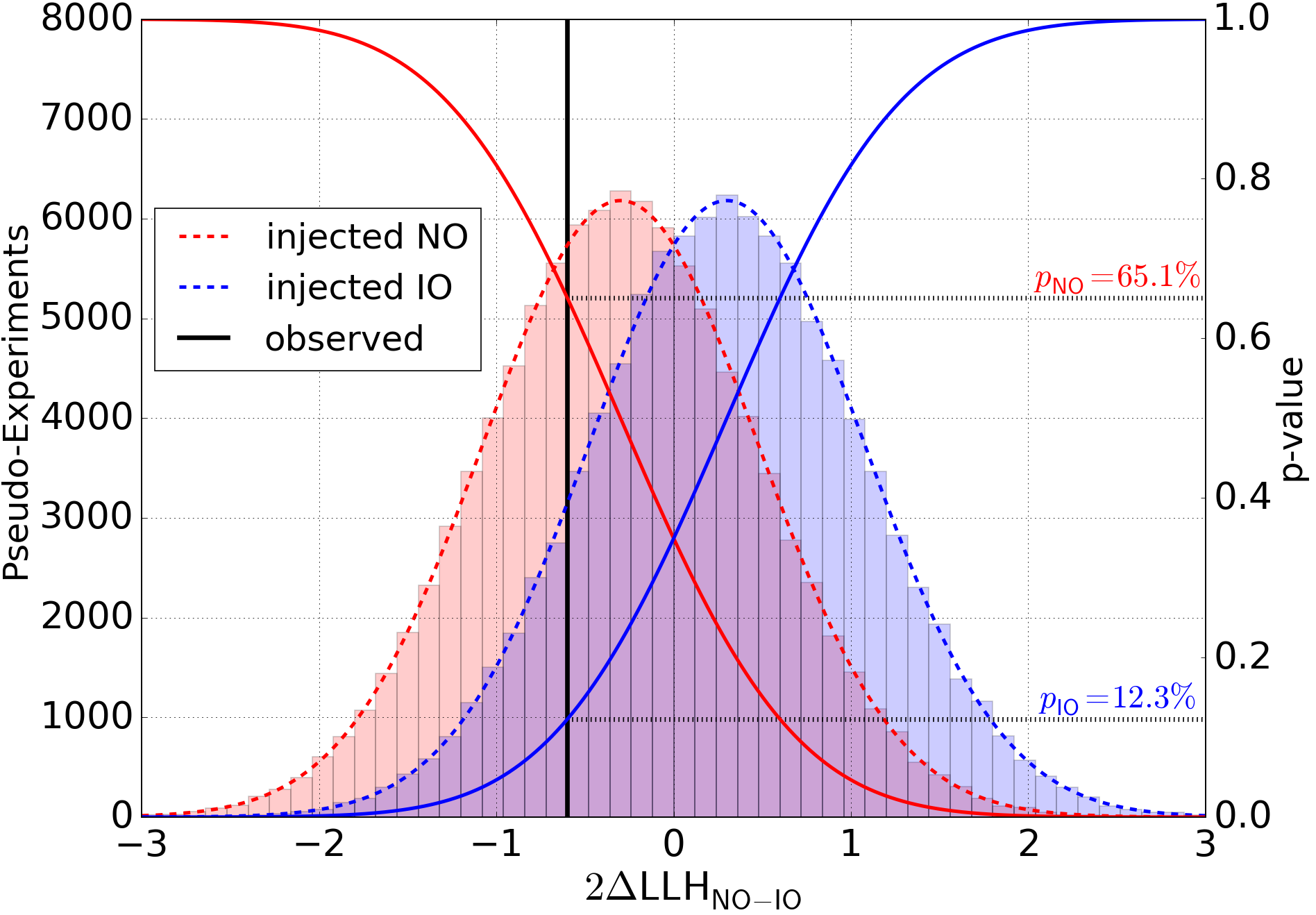}
  \end{center}
  \caption{\newtext{Sketch of the frequentist method, using idealized distributions to illustrate the concepts. 
  The red (blue) distribution will be derived from from PTs assuming the $\mathcal{H}_\mathrm{NO}$ ($\mathcal{H}_\mathrm{IO}$) hypothesis.} The black vertical line represents a hypothetically observed value of $\Delta \mathrm{LLH}_\mathrm{NO-IO}$. The resulting $p$-values (right, vertical axis) for the hypotheses are derived from the cumulative density distributions, marked as red (blue) solid lines for NO (IO)}
  \label{fig:SketchPseudos}
\end{figure}

This process of creating PTs for $\eta^{\mathrm{NO}}$ and $\eta^{\mathrm{IO}}$ and fitting them with both hypotheses is repeated several times to estimate a TS distribution for each of the ordering hypotheses.
The TS distributions for NO and IO are then used to estimate the analysis sensitivity, {\it i.e.}\ the expected $p$-values for the exclusion of each hypothesis. To do this, the fraction of PTs for NO (IO) that is to the right (left) of the median of the IO (NO) distribution is taken as the expected $p$-value for the exclusion of the NO (IO) hypothesis, if IO (NO) is the true ordering. This is sketched in Figure~\ref{fig:SketchPseudos} for two generic distributions.

\begin{figure}[tb]
  \begin{center}
    \includegraphics[width=0.48\textwidth]{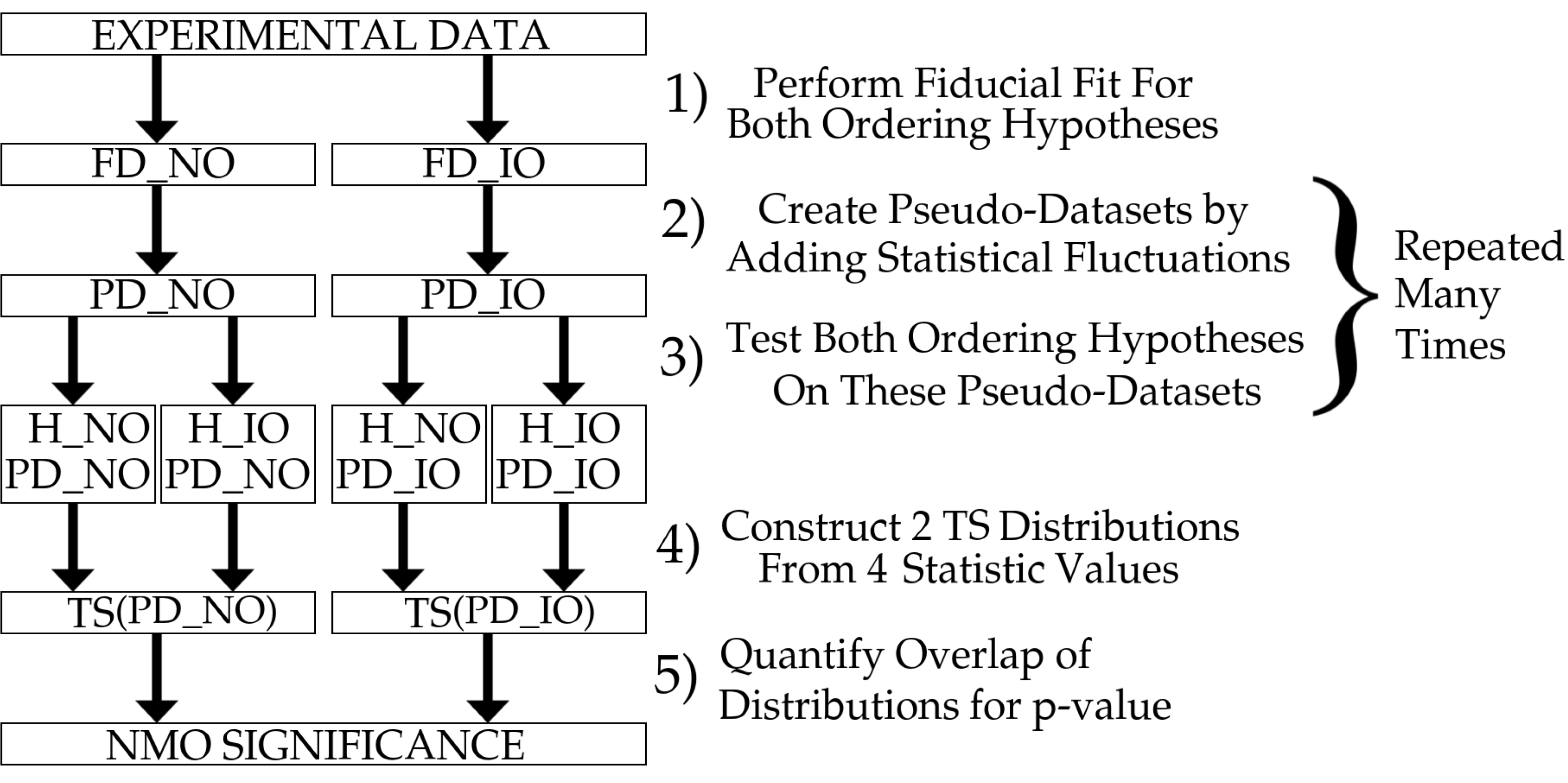} 
  \end{center}
  \caption{
  Flow-chart representing the procedure of the frequentist method used to derive $p$-values for the NO and IO hypotheses. \newtext{Abbreviations, as defined in the text, are FD (fiducial fits), PD (pseudo-dataset), H (hypothesis), and TS (test statistic).}}
  \label{fig:FCFlowchart}
\end{figure}

The frequentist method is summarized as a flow-chart in Figure~\ref{fig:FCFlowchart}. Note that this procedure is similar to the treatment of data, described in Section~\ref{sec:firstresults}, where the experimental fit is used as fiducial fit to produce PTs.
Unfortunately, the frequentist method is computationally very expensive. Thus, for performing more detailed parameter studies, a second, faster method is used. 

The second method for deriving sensitivities is an \textit{Asimov approach} adapted from~\cite{Ciuffoli:2013rza}. Instead of generating PTs, the total MC template, with no Poissonian fluctuations, is fitted directly for both hypotheses. In the following, we refer to this MC template as the generated-ordering (GO) hypothesis, $\mathcal{H}_\mathrm{GO}$, where the GO can be either NO or IO. This is then fitted under assumptions of both hypotheses, NO and IO, where the hypothesis used in the fit is called the fitted-ordering (FO) hypothesis, $\mathcal{H}_\mathrm{FO}$. The negative log-likelihood value obtained from the fit is $\overline{\mathrm{LLH}}_\mathrm{FO}(\mathcal{H}_\mathrm{GO}) = 0$ if $\mathcal{H}_\mathrm{FO} = \mathcal{H}_\mathrm{GO}$ and $\overline{\mathrm{LLH}}_\mathrm{FO}(\mathcal{H}_\mathrm{GO}) > 0$ otherwise, where the bars indicate that the values were obtained by injecting the template of the GO directly.


The resulting value of $2\overline{\Delta\mathrm{LLH}}_\mathrm{NO-IO}$ is assumed to be 
representative for the behavior obtained using PTs.
The sensitivity to the generated ordering, $n_\sigma^\mathrm{GO} $, in terms of one-sided Gaussian standard deviations is
\begin{equation} \label{eq:SensAsimov}
n^\mathrm{GO}_{\sigma}  = \frac{\overline{\Delta \mathrm{LLH}}_\mathrm{NO-IO}(\mathcal{H}_\mathrm{GO}) - \overline{\Delta\mathrm{LLH}}_\mathrm{NO-IO}(\mathcal{H}_{\widetilde{\mathrm{GO}}}) }{ \sqrt{ 2\overline{\Delta\mathrm{LLH}}_\mathrm{NO-IO}(\mathcal{H}_{\widetilde{\mathrm{GO}}}) } },
\end{equation}
where $\widetilde{GO} \in \{ \mathrm{IO},\mathrm{NO} \}$ is the opposite hypothesis to GO, generated with the best-fit set of systematic parameters $ \eta^{\widetilde{\mathrm{GO}}} \in \{ \eta^{\mathrm{IO}}, \eta^{\mathrm{NO}} \}$ 
corresponding to the set $ \eta^{\mathrm{GO}}\in \{\eta^{\mathrm{NO}}, \eta^{\mathrm{IO}} \}$ used for $\mathcal{H}_\mathrm{GO}$. 
Note that the sensitivity $n_\sigma^\mathrm{GO} $ describes the expected $p$-value for the exclusion of the $\widetilde{\mathrm{GO}} $ hypothesis in the case that the true ordering is the GO\,\cite{Ciuffoli:2013rza}.


The choice of one- instead of two-sided Gaussian standard deviations is motivated by the fact that an experiment with no sensitivity to the NMO, {\it i.e.}\ if the two distributions for NO and IO in Figure~\ref{fig:SketchPseudos} were identical,  would lead to a $50\%$ chance of obtaining the correct ordering by random chance. This should not be misinterpreted as sensitivity and thus should give $n_\sigma^\mathrm{NO,\,IO} = 0$, which is the case for one-sided but not two-sided Gaussians.

\begin{figure}[tb]
  \begin{center}
    \includegraphics[width=0.48\textwidth]{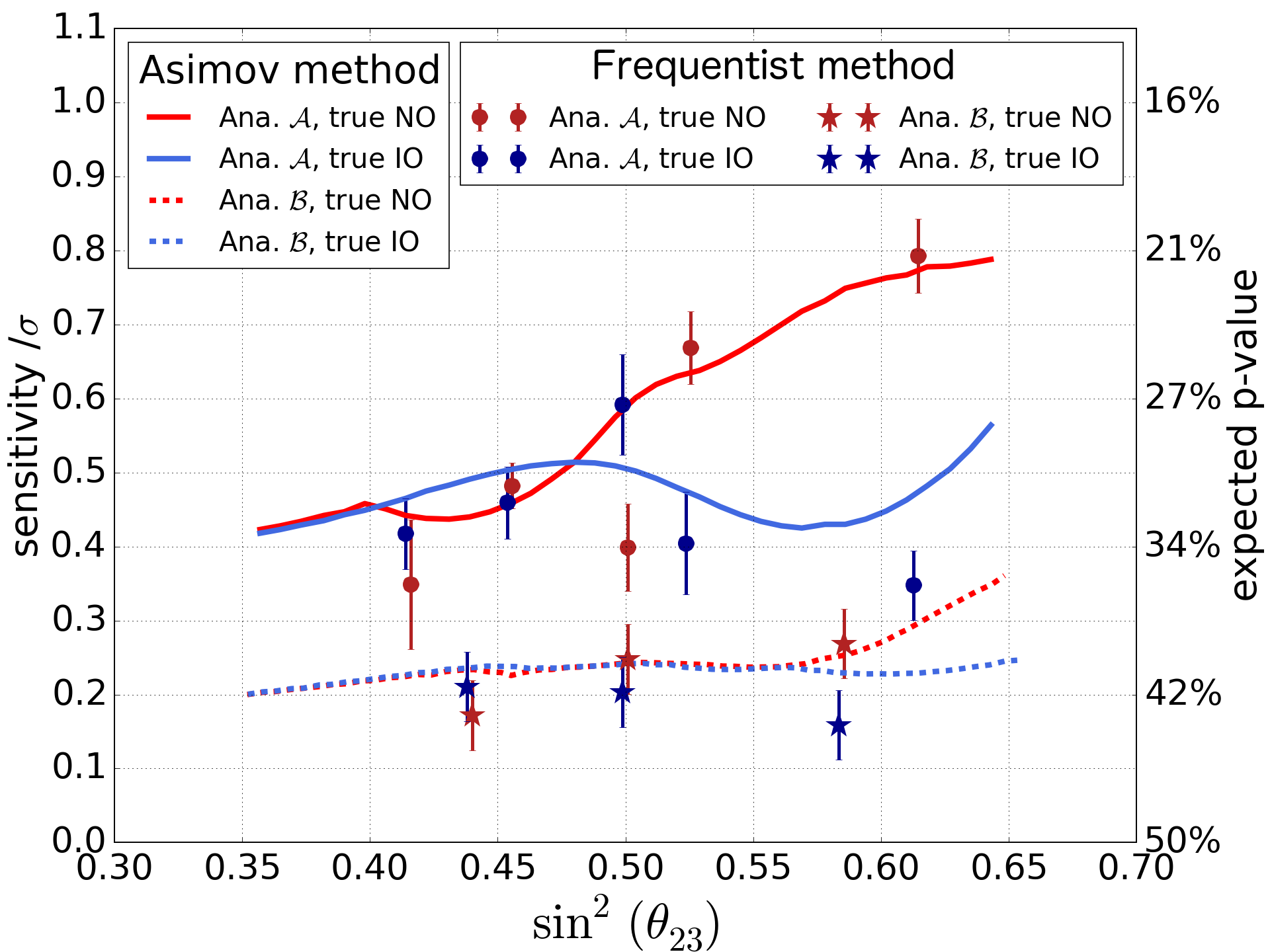} 
  \end{center}
  \caption{Sensitivities of Analyses~\greco and~\dragon to the NMO in terms of one-sided Gaussian sigmas (left vertical axis) and $p$-values (right vertical axis) derived by the Asimov-method (lines), and validated at certain values of $\sin^2(\theta_{23})$ using the frequentist method (markers). \newnewtext{The statistical errors on the frequentist points arise from the finite number of PTs used due to the computationally intensive nature of the frequentist method.}}
  \label{fig:Sensitivity}
\end{figure}

The resulting sensitivities for both methods are shown in Figure~\ref{fig:Sensitivity}, as a function of the true value of $\sin^2(\theta_{23})$.
The blue and red lines indicate the result from the Asimov method for Analysis~\greco\ (solid lines) and Analysis~\dragon\ (dashed lines). 
The sensitivities are validated at certain values of $\sin^2(\theta_{23})$ using the frequentist method, as indicated by the circle (\greco) and star (\dragon) markers, where the uncertainties arise from the finite number of PTs.


As visible in Figure~\ref{fig:Sensitivity}, the resulting sensitivity is $< 1\sigma$ for both orderings and analyses. Moreover, Analysis~\greco\ is more sensitive to the NMO than Analysis~\dragon, which is due to the increased statistics, the additional bins in PID, energy and zenith, and the reduced impact from limited Monte Carlo statistics, due to the usage of KDEs in the generation of Monte Carlo templates.

Note that a characteristic shape is found for the $\sin^{2}(\theta_{23})$-dependence of $n_\sigma^\mathrm{NO,\,IO}$, which is different for the NO and IO hypotheses. The observed features are similar to those found for the PINGU sensitivity in~\cite{ref:PINGULoI}. They arise from the interplay of the two independent octant fits for $\overline{\mathrm{LLH}}_\mathrm{GO}$ and $\overline{\mathrm{LLH}}_{\widetilde{\mathrm{GO}}}$, 
used to calculate the values of $\overline{\Delta\mathrm{LLH}}_\mathrm{NO-IO}(\mathcal{H}_\mathrm{GO})$ and $\overline{\Delta\mathrm{LLH}}_\mathrm{NO-IO}(\mathcal{H}_{\widetilde{\mathrm{GO}}})$ in Equation~\ref{eq:SensAsimov}, where the preferred octant is not necessarily the true one in the case that $\widetilde{\mathrm{GO}}$ is fitted. As a result, the behavior of $n_\sigma^\mathrm{NO,\,IO}$ changes each time the octant is flipped for one of the two negative log-likelihood differences ($\Delta\mathrm{LLH}$) in Equation~\ref{eq:SensAsimov}.

The observed sensitivities for the Asimov method agree roughly with the PTs, while perfect agreement is not expected due to several approximations used in the Asimov-method\,\cite{Ciuffoli:2013rza}. However, the Asimov method is used as an estimator for the true sensitivity.

Note that for some observed values in Figure~\ref{fig:SketchPseudos} the $p$-values for both hypotheses can be small, in case the observed data agrees with neither the NO nor IO hypotheses. For example, this could be the case for $\Delta \mathrm{LLH}_\mathrm{NO-IO}> 2$ or $\Delta \mathrm{LLH}_\mathrm{NO-IO}< -2$, which is in the tail of both distributions in Figure~\ref{fig:SketchPseudos}. In this case, the small $p$-value might lead to the wrong impression that the data clearly favors the alternative over the null hypothesis.
To properly account for this, the \(p\)-values are combined into a  \(\mathrm{CL}_S\)-value,
\begin{equation} \label{eqn:CLs}
  \mathrm{CL}^{\greco/\dragon}_S\left(\mathcal{H}_\mathrm{TO}\right)=\frac{p_{\greco/\dragon}(\mathcal{H}_\mathrm{TO})}{1-p_{\greco/\dragon}(\mathcal{H}_{\widetilde{\mathrm{TO}}})}\,,
\end{equation}
\noindent where TO is the tested ordering and $\widetilde{\mathrm{TO}}$ is the opposite ordering hypotheses.
This equation is taken from~\cite{Qian:2014nha} where a more detailed discussion of its derivation can be found. Its value is limited to $\mathrm{CL}_S \in [0,1]$, where $\mathrm{CL}_S \approx 1 $ indicates no preference for one hypothesis over the other and $\mathrm{CL}_S \approx 0 $ indicates a strong disfavoring of the given hypothesis. The $\mathrm{CL}_S$ value can be interpreted as \textit{confidence} in the result with a confidence level of $1-\mathrm{CL}_S$. More illustratively, the $\mathrm{CL}_S$ value describes how much less likely the observed value would occur under the disfavored hypothesis, compared to the favored one.

Finally, potential improvements of the sensitivity are tested for Analysis~\greco.  By fixing individual and combinations of systematic parameters in the Asimov fit, the absolute gain in sensitivity from an improved understanding of systematic uncertainties is found to be small, except for the oscillation parameters.
This is due to the weak NMO signature, which barely pulls the systematic parameters and thus is only weakly affected by fixing them. Instead, it is found that the sensitivity could be improved in the future by additional data statistics and improvements on the event reconstruction, which reduce the smearing-out of the NMO signature due to the low resolution in energy, zenith, and PID at the lowest energies\,\cite{LeuermannThesis}.

\section{Results}
\label{sec:firstresults}

\begin{figure}
    \centering
    \includegraphics[width=\columnwidth]{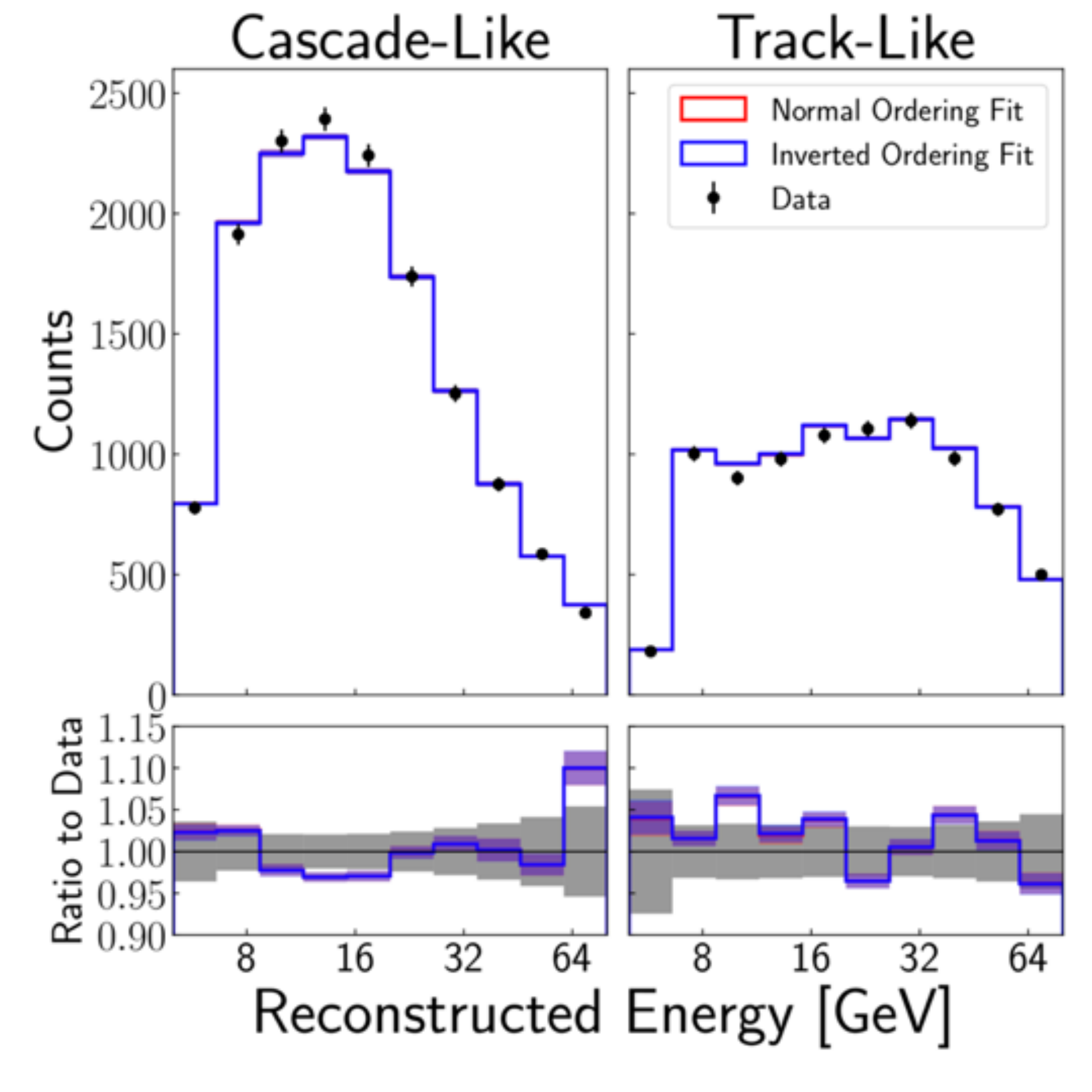}
    \includegraphics[width=\columnwidth]{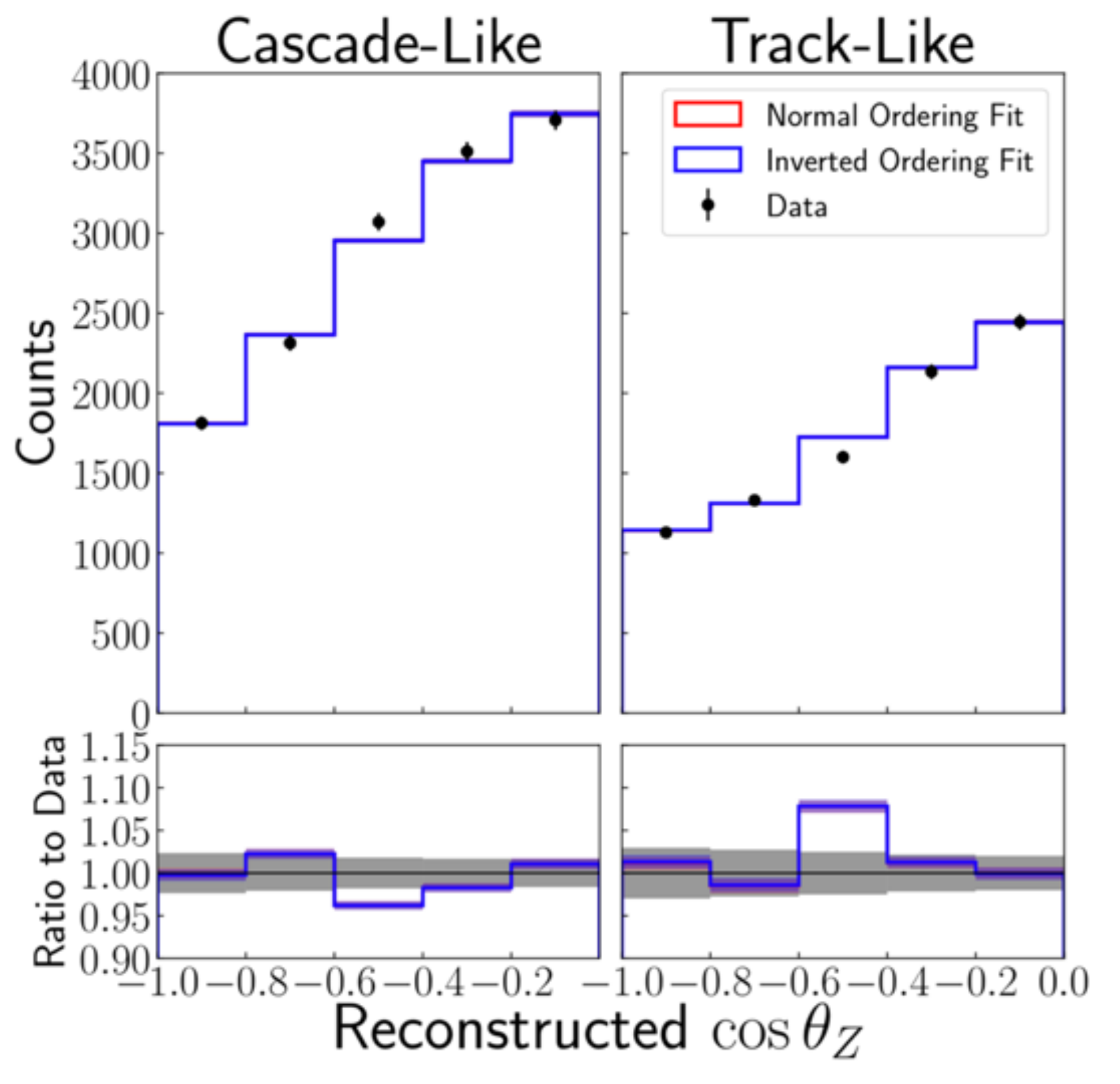}
    \caption{\newtext{The energy and zenith-angle distributions of the data from Analysis~\dragon. Also shown are the best-fit simulations for both orderings, where the red and blue lines fall almost on top of each other.}}
    \label{fig:DragonSpectra}
\end{figure}

For both analyses, the experimental data is fitted with the likelihood method, described in Section~\ref{sec:Analyses}.
\newtext{The data, along with the best-fit predictions, are shown for Analysis~\dragon in Figure~\ref{fig:DragonSpectra}.}
The resulting best-fit values for all systematic parameters are shown in Table~\ref{tab:systematics}. The observed pulls are within the expected ranges for all parameters, taking statistical fluctuations and the uncertainties of the true value of each parameter into account. The corresponding values of the metric for the NO (IO) hypothesis are $2\mathrm{LLH} = 293.38$ ($294.12$) for Analysis~\greco\ and  $\chi^2=107.82$ ($107.50$) for Analysis~\dragon. The metric is used as a \textit{goodness-of-fit} estimator for the agreement of data and Monte Carlo by comparing these values to the expectation from PTs. The resulting $p$-values for Analyses~\greco\ and~\dragon\ are  $p_\mathrm{gof}^\greco=43.5\%$ and $p_\mathrm{gof}^\dragon=11.0\%$, indicating the data to be well-described by the MC templates. 

For Analyses~\greco\ and~\dragon, the observed values of the test-statistic are  $2\Delta\mathrm{LLH}_\mathrm{NO-IO}=-0.738$ and $\Delta\chi^2_\mathrm{NO-IO}=0.3196$. Thus, the fits for the main result (\greco) and the confirmatory result (\dragon) prefer NO and IO, respectively, \newnewtext{while both results are compatible within their statistical uncertainties, i.e. both results have a test statistic within one unit of zero.}

\begin{figure}[tb]
  \begin{center}
\includegraphics[width=0.48\textwidth]{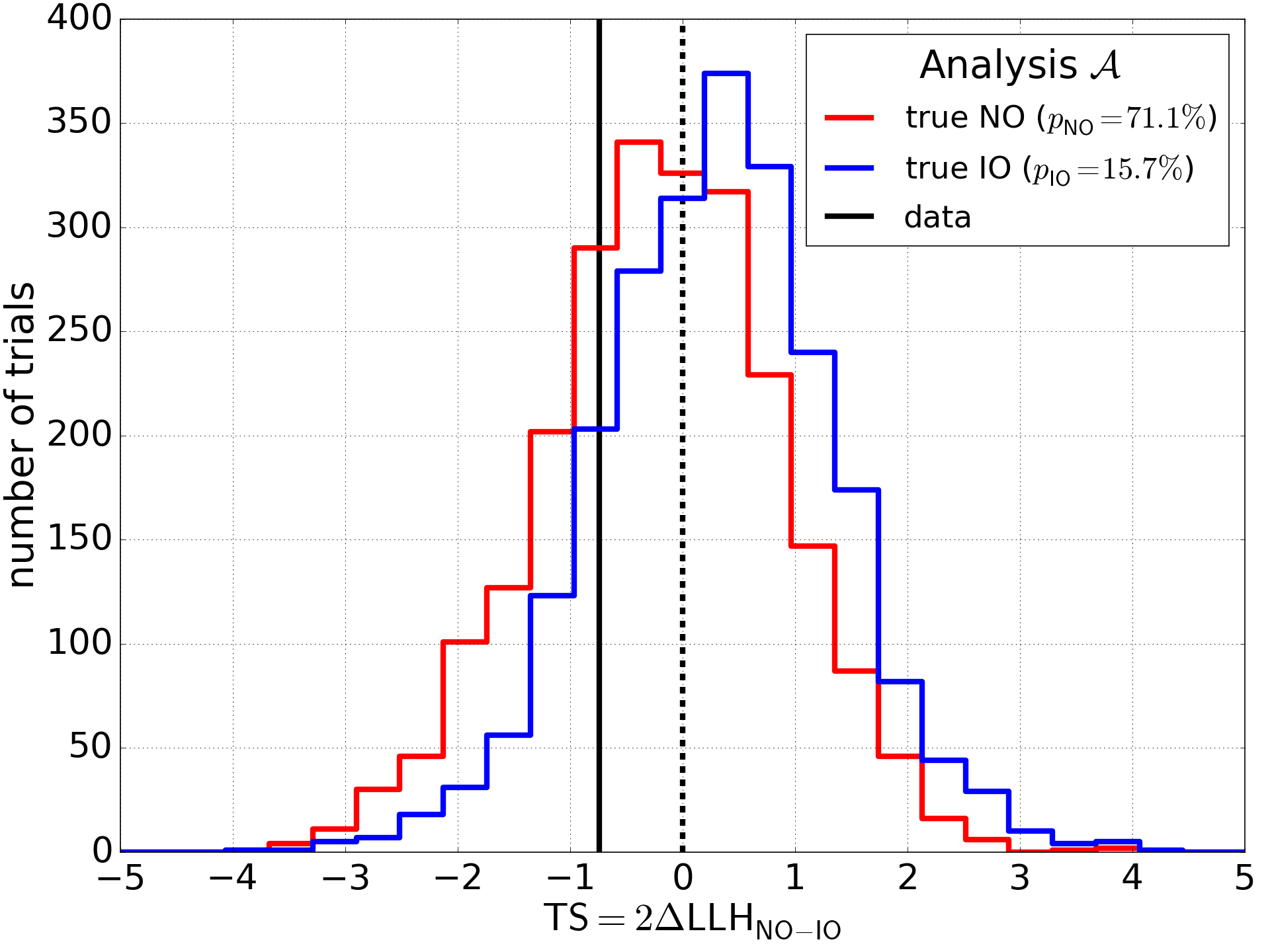} 
\\
    \includegraphics[width=0.48\textwidth]{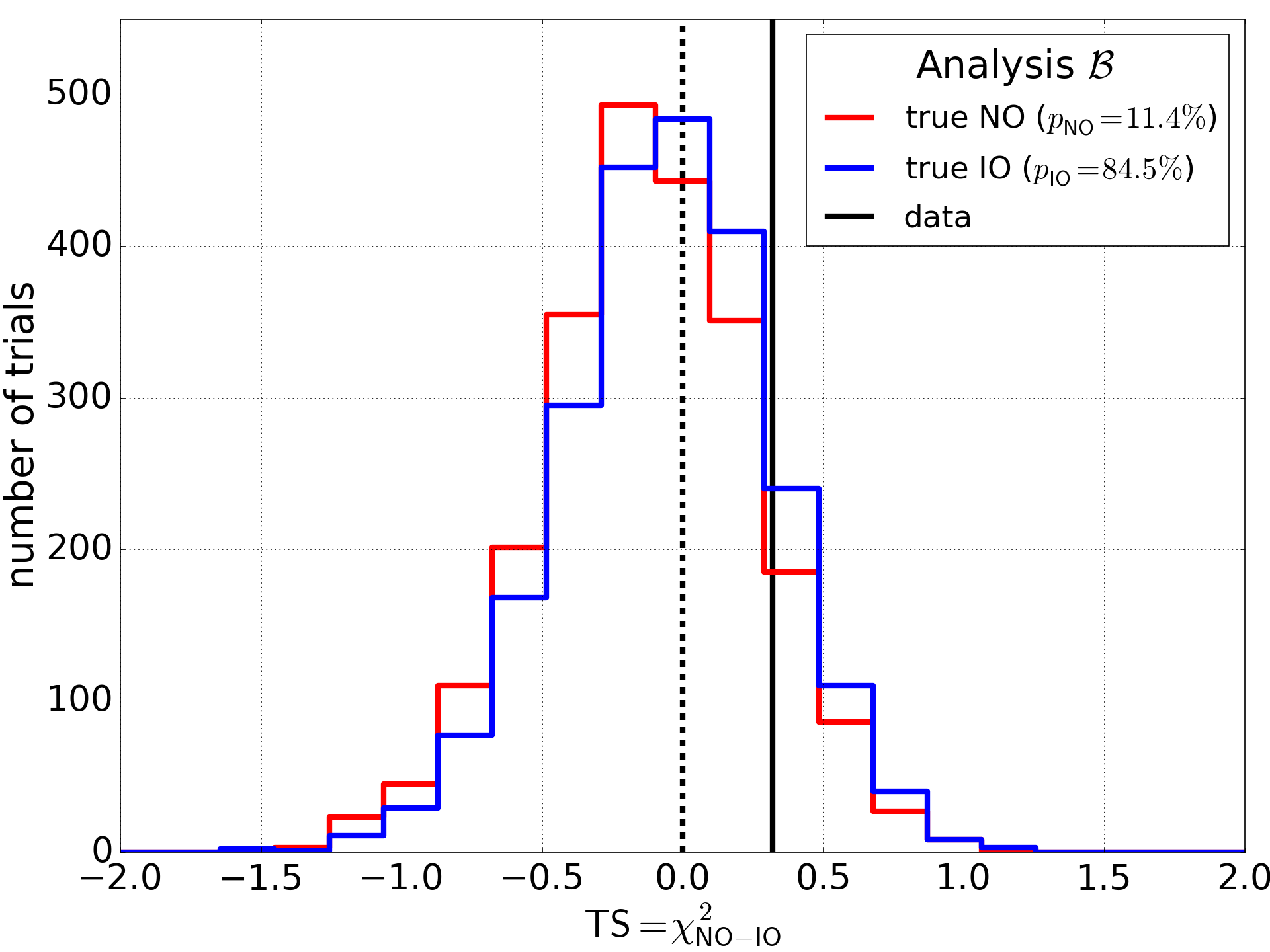}
  \end{center}
  \caption{Distribution of the TS from PTs, generated with the best-fit systematic parameters $\eta^{\mathrm{NO}}$ and $\eta^{\mathrm{IO}}$ from Table~\ref{tab:Params} for Analysis~\greco\ (top) and Analysis~\dragon\ (bottom). The red and blue distributions are obtained for the NO and IO hypotheses, respectively, while the black, solid vertical line shows the observed value in data, giving the $p$-values for the NO and IO hypotheses stated in the legends.}
  \label{fig:GRECOTSDists}
\end{figure}

To estimate the corresponding $p$-values, PTs are generated with the best-fit parameters $\eta^\mathrm{NO}$ and $\eta^\mathrm{IO}$ from Table~\ref{tab:systematics}; for each PT, both ordering hypotheses are fitted. The resulting distributions of $\mathrm{TS} = 2\Delta\mathrm{LLH}_\mathrm{NO-IO}$ and $\mathrm{TS} = \chi^2_\mathrm{NO-IO}$ are shown in Figure~\ref{fig:GRECOTSDists}. The experimentally observed value is indicated by the solid, vertical black line, indicating the preference for  Normal over Inverted Ordering in Analysis~\greco\ and  Inverted over Normal Ordering in Analysis~\dragon.

The resulting $p$- and $\mathrm{CL}_S$-values for the main result are 
\begin{alignat}{3}
p_\greco(\mathcal{H}_\mathrm{NO}) &= 71.1\% && ~(\mathrm{CL}^\greco_{S}(\mathcal{H}_\mathrm{NO})&&=83.0\%), \\
p_\greco(\mathcal{H}_\mathrm{IO})\, &= 15.7\% && ~(\mathrm{CL}^\greco_S(\mathcal{H}_\mathrm{IO})&&=53.3\%), 
\end{alignat}
while for the confirmatory result we find
\begin{alignat}{3}
p_\dragon(\mathcal{H}_\mathrm{NO}) &= 11.4\% && ~(\mathrm{CL}^\dragon_S(\mathcal{H}_\mathrm{NO})&&=73.5\%), \\
p_\dragon(\mathcal{H}_\mathrm{IO})\, &= 84.5\% && ~(\mathrm{CL}^\dragon_S(\mathcal{H}_\mathrm{IO})&&=95.4\%).
\end{alignat}

\begin{figure}[tb]
  \begin{center}
    \includegraphics[width=0.48\textwidth]{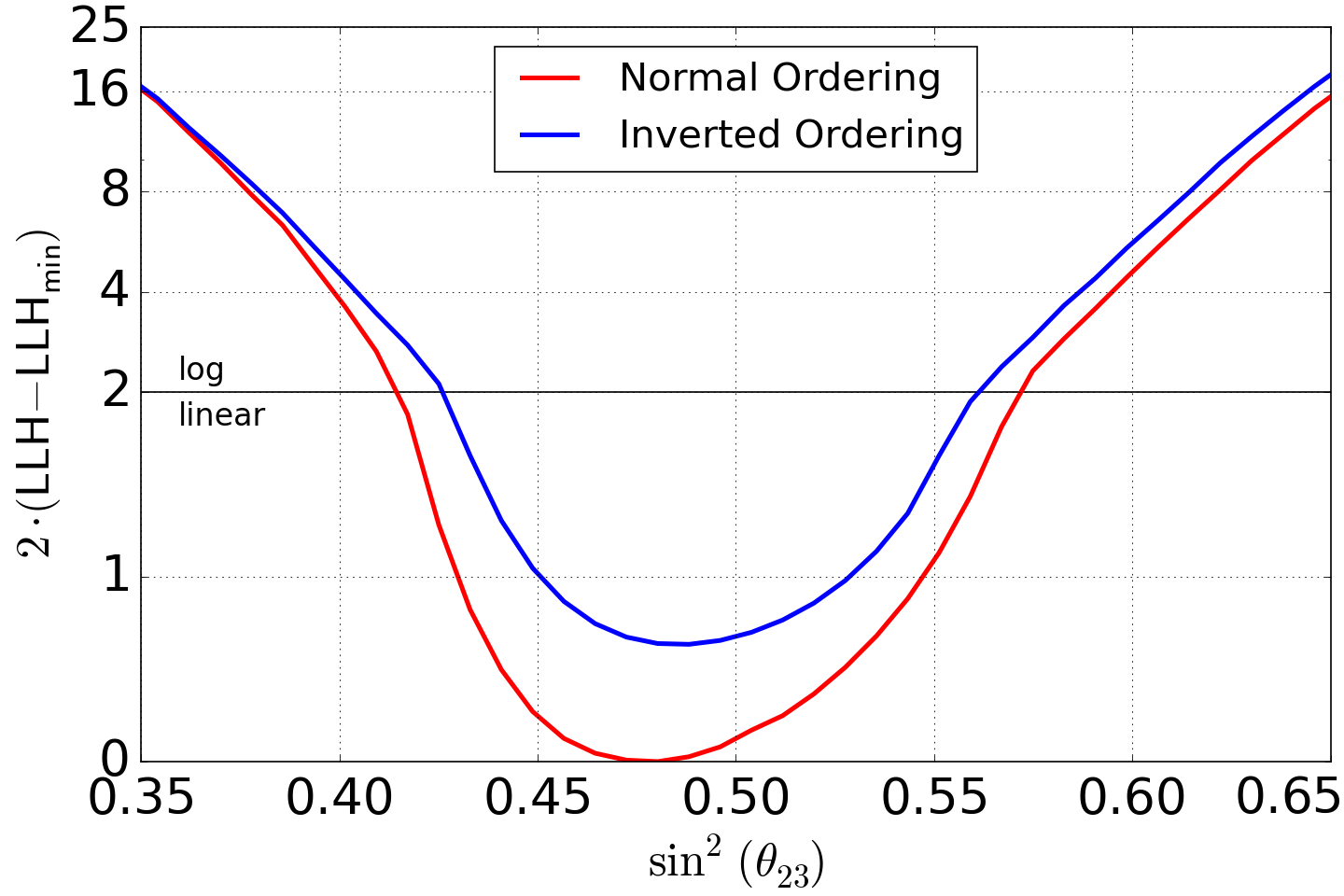}
  \end{center}
  \caption{The negative log-likelihood (LLH) as a function of $\sin^2(\theta_{23})$ for Analysis~\greco, relative to the global minimum $\mathrm{LLH}_\mathrm{min}$. The preference for NO over IO is visible over all the range of $\sin^2(\theta_{23})$ with the best-fit for both orderings being in the lower octant ($\sin^2(\theta_{23}) < 0.5$).}
  \label{fig:1DContour}
\end{figure}

In addition to testing the NMO with PTs, the likelihood is scanned across $\sin^2(\theta_{23})$ for the more sensitive Analysis~\greco\ and both ordering hypotheses. The resulting scan is shown in Figure~\ref{fig:1DContour}, where the LLH is shown with respect to its global minimum. The vertical offset between the NO and IO curves indicates the preference for NO over IO, which is visible at all values of $\sin^2(\theta_{23})$. The observed minimum is in the lower octant, near $\sin^2(\theta_{23})= 0.455$, for both orderings, while maximal mixing is separated from the best-fit point by only $2\Delta \mathrm{LLH}_\mathrm{NO-IO} = 0.128$ ($0.681$) for NO (IO). As a result, the preference for the lower octant is small, such that a substantial range of $\sin^2(\theta_{23})>0.5$ is still compatible with the observed data for NO and IO.

Note that the preference for NO over IO in Analysis~\greco\ already indicates an observed preference for matter effects in data (cf. Section~\ref{sec:intro}), i.e. a preference for matter effects over vacuum oscillations. 
To quantify this preference, the fit is repeated assuming vacuum oscillations. The resulting log-likelihood difference between matter effects (MA) and vacuum oscillations (VA) is $\Delta \mathrm{LLH}_\mathrm{MA-VA} = -0.869$ ($-0.500$) in case NO (IO) is assumed. Thus, matter effects are preferred over vacuum oscillations, independent of the assumption on the NMO.
\newtext{The $p$-values and CL$_{S}$-values that quantify the preference for matter effects (Mat) and vacuum oscillations (Vac) are}
\newcolour
\begin{eqnarray}
p(\mathcal{H}_{\mathrm{Mat}}|\mathcal{H}_{\mathrm{NO}}) &= 62.3\%, \quad \mathrm{CL}_{S}(\mathcal{H}_{\mathrm{Mat}}|\mathcal{H}_{\mathrm{NO}}) &= 71.0\%,\\
p(\mathcal{H}_{\mathrm{Vac}}|\mathcal{H}_{\mathrm{NO}}) &= 12.3\%, \quad \mathrm{CL}_{S}(\mathcal{H}_{\mathrm{Vac}}|\mathcal{H}_{\mathrm{NO}}) &= 32.6\%,\\
p(\mathcal{H}_{\mathrm{Mat}}|\mathcal{H}_{\mathrm{IO}}) &= 53.2\%, \quad \mathrm{CL}_{S}(\mathcal{H}_{\mathrm{Mat}}|\mathcal{H}_{\mathrm{IO}}) &= 68.4\%,\\
p(\mathcal{H}_{\mathrm{Vac}}|\mathcal{H}_{\mathrm{IO}}) &= 22.2\%, \quad \mathrm{CL}_{S}(\mathcal{H}_{\mathrm{Vac}}|\mathcal{H}_{\mathrm{IO}}) &= 47.4\%.
\end{eqnarray}
\oldcolour

\section{Conclusion}
\label{sec:conclusion}


We have
\newtext{developed two independent likelihood analyses to demonstrate the extraction of the neutrino mass ordering from atmospheric neutrino data. We have applied these analyses to three years of IceCube DeepCore data. The first analysis aims}
for an optimized sensitivity with DeepCore, the second for an analysis chain as similar as possible to the proposed NMO analysis with PINGU\,\cite{ref:PINGULoI}. The sensitivities were estimated with two independent methods. For the more sensitive, main analysis, the sensitivity was found to be $\sim 0.45-0.65\,\sigma$ (one-sided Gaussian), within the most interesting region close to maximum mixing ($\sin^2(\theta_{23}) \in \left[0.45, 0.55\right]$) for both orderings, while for the confirmatory analysis, the sensitivity was found to be $\sim 50\%$ smaller.

Due to the weak signature of the NMO in DeepCore, the sensitivity is found to be mostly unaffected by improvements in the understanding of systematic uncertainties. Instead, a future gain in sensitivity might come from additional statistics or potential improvements in the resolution of the event reconstruction.

\newtext{The analyses presented here find the data to be fully compatible with both mass orderings.}
The main analysis observes a preference for NO over IO at  $2\Delta\mathrm{LLH}_\mathrm{NO-IO} =-0.738$, which corresponds to a $p$-value of $15.3\%$ ($\mathrm{CL}_\mathrm{s}=53.3\%$) for the IO hypothesis, based on the presented frequentist method.
This result is in line with recently reported preferences for the NO by Super-Kamiokande\,\cite{SuperK_newest}, T2K\,\cite{T2K_newest}, NO$\nu$A\,\cite{NOVA_newest}, MINOS\,\cite{MINOS_newest}, and recent global best fits\,\cite{NuFit_Web, NuFit_Paper}. However, it complements these results due to the higher energy range used for determining the NMO ($E_\nu \gtrsim 5\,\mathrm{GeV}$) and the fact that it is independent of the value of $\delta_\mathrm{CP}$. 
Finally, the data indicates a preference for matter effects over vacuum oscillations, independent of the assumption on the NMO.
 

\newtext{
The study presented here allows us to consider what future steps will allow a determination of the NMO with atmospheric neutrino data. Given the statistically-limited nature of this result, it is clear that a reduction of systematic uncertainties is not a priority, and we have performed studies to show that even the most optimistic reduction of systematic uncertainties can achieve at most a 10\% improvement in the NMO sensitivity of this dataset\cite{LeuermannThesis}. The same study also showed that a removal of backgrounds (atmospheric muons and triggered noise) delivers at most a 5\% improvement in sensitivity. In the coming years, a factor of four more statistics is expected from DeepCore (including both additional data and expected data-selection improvements), and this can result in a factor of two improvement in sensitivity. A more significant improvement that can be made is in the measurement resolutions: our studies \cite{LeuermannThesis} show that a 50\% improvement in resolution on both neutrino direction and $\log_{10}(E_{\nu})$ would produce a factor of two improvement in the sensitivity of this dataset. To achieve an NMO determination in a reasonable timescale, a final necessary improvement is a lowering of the neutrino energy threshold; this, along with the improved resolutions, can be achieved by the PINGU concept~\cite{ref:PINGUVision,ref:PINGULoI} that reduces the energy threshold to below 10\,GeV to enable a $3\sigma$ determination of the NMO for even the least optimistic values of the oscillation parameters.
}

Besides the experimental result, the presented analyses provide a \textit{proof-of-concept} for determining the NMO from matter effects in atmospheric neutrino oscillations with the IceCube Upgrade\,\cite{JP_Gen2Phase1_NutauApp} or PINGU\,\cite{ref:PINGULoI}. 
They test the full analysis chain by means of real DeepCore data and validate the understanding and treatment of systematic uncertainties, which are largely consistent with those 
that will be encountered by future IceCube extensions.

\begin{acknowledgements}

USA -- U.S. National Science Foundation-Office of Polar Programs,
U.S. National Science Foundation-Physics Division,
Wisconsin Alumni Research Foundation,
Center for High Throughput Computing (CHTC) at the University of Wisconsin-Madison,
Open Science Grid (OSG),
Extreme Science and Engineering Discovery Environment (XSEDE),
U.S. Department of Energy-National Energy Research Scientific Computing Center,
Particle astrophysics research computing center at the University of Maryland,
Institute for Cyber-Enabled Research at Michigan State University,
and Astroparticle physics computational facility at Marquette University;
Belgium -- Funds for Scientific Research (FRS-FNRS and FWO),
FWO Odysseus and Big Science programmes,
and Belgian Federal Science Policy Office (Belspo);
Germany -- Bundesministerium f\"ur Bildung und Forschung (BMBF),
Deutsche Forschungsgemeinschaft (DFG),
Helmholtz Alliance for Astroparticle Physics (HAP),
Initiative and Networking Fund of the Helm\-holtz Association,
Deutsches Elektronen Synchrotron (DESY),
and High Performance Computing cluster of the RWTH Aachen;
Sweden -- Swedish Research Council,
Swedish Polar Research Secretariat,
Swedish National Infrastructure for Computing (SNIC),
and Knut and Alice Wallenberg Foundation;
Australia -- Australian Research Council;
Canada -- Natural Sciences and Engineering Research Council of Canada,
Calcul Qu\'ebec, Compute Ontario, Canada Foundation for Innovation, WestGrid, and Compute Canada;
Denmark -- Villum Fonden, Danish National Research Foundation (DNRF), Carlsberg Foundation;
New Zealand -- Marsden Fund;
Japan -- Japan Society for Promotion of Science (JSPS)
and Institute for Global Prominent Research (IGPR) of Chiba University;
Korea -- National Research Foundation of Korea (NRF);
Switzerland -- Swiss National Science Foundation (SNSF);
United Kingdom -- Science and Technology Facilities Council (STFC), part of UK Research and Innovation.
The IceCube collaboration acknowledges the significant contributions to this manuscript from Martin Leuermann and Steven Wren.

\end{acknowledgements}

\bibliographystyle{epjc}
\bibliography{main.bib}

\end{document}